\documentclass[a4paper, 12pt]{article}

\usepackage[a4paper,margin=1in]{geometry}
\usepackage[english]{babel}
\usepackage[utf8x]{inputenc}
\usepackage[T1]{fontenc}
\usepackage{amsmath, amssymb, bm, bbm}
\usepackage{setspace}
\usepackage{natbib}
\usepackage{graphicx}
\usepackage[colorinlistoftodos]{todonotes}
\usepackage[colorlinks=true, allcolors=blue]{hyperref}

\providecommand{\keywords}[1]{\textbf{Keywords:} #1}
\DeclareMathOperator*{\argmin}{arg\,min}

\title{Bayesian spatial clustering of extremal behaviour for hydrological variables}
\author{Christian Rohrbeck and Jonathan A.\ Tawn\vspace{0.3cm}\\
{\small Department of Mathematics and Statistics, Lancaster University}}
\date{\vspace{-0.6cm}}

\begin{document}
\maketitle

\doublespacing

\begin{abstract}~\vspace{-0.5cm}\\
To address the need for efficient inference for a range of hydrological extreme value problems, spatial pooling of information is the standard approach for marginal tail estimation. We propose the first  extreme value spatial clustering methods which account for both the similarity of the marginal tails and the spatial dependence structure of the data to determine the appropriate level of pooling. Spatial dependence is incorporated in two ways: to determine the cluster selection and to account for dependence of the data over sites within a cluster when making the marginal inference. We introduce a statistical model for the pairwise extremal dependence which incorporates distance between sites, and accommodates our belief that sites within the same cluster tend to exhibit a higher degree of dependence than sites in different clusters. We use a Bayesian framework which learns about both the number of clusters and their spatial structure, and that enables the inference of site-specific marginal distributions of extremes to incorporate uncertainty in the clustering allocation. The approach is illustrated using simulations, the analysis of daily precipitation levels in Norway and daily river flow levels in the UK.
\end{abstract}

~\vspace{-0.6cm}\\
\keywords{ Bayesian clustering; Extreme value analysis; Spatio-temporal modelling;

~\hspace{1.8cm}Reversible jump Markov chain Monte Carlo.}

\section{Introduction}
Statistical models for estimating the frequency and size of flood and severe rainfall events are required by decision makers to construct effective protection measures and by risk analysts to set insurance premiums. Since such extreme events occur rarely at a site of interest, model-based estimates for the behaviour of extremes at a site are usually derived using a small number of observations, inducing high uncertainty. With a view to obtaining more reliable estimates, pooling of information from other sites can be used; this is the basis of regional methods which are widely used for environmental, meteorological or hydrological hazards \citep{Hosking1993, Casson1999, Sang2009, Asadi2018}. Natural questions that arise in this context are what criteria should be used to group sites, and how should the uncertainty in the clustering allocation, given this selected criterion, be reflected in the uncertainty in the estimated site-specific marginal distributions, and what do the estimated clusters look like?

We propose a novel approach for spatial clustering of extremes and the subsequent inference which coherently addresses these questions. Our methodology is motivated by two applications: modelling weather-related insurance claims and flood risk analysis described in Section~\ref{sec:Data}. In both applications, we are interested in the extremal behaviour of a hydrological variable across multiple spatial locations. Each of these problems requires the marginal analysis of extreme values at different locations, whilst accounting for spatial structure both in the marginal distributions and in the dependence of the data from across sites. 

Our approach for modelling both marginal and dependence structures is to use statistical models that have been asymptotically justified by extreme value theory  \citep{Coles2001,Beirlant2004}.  When considering the extremes of a univariate random variable $X$, we adopt the peaks-over threshold approach. Exceedances by $X$ of a high threshold $u$ are then modelled using the generalized Pareto distribution GPD$(\psi,\nu)$ with
\begin{equation}
\mathbb{P}(X\leq x+u\mid X>u) ~=~ 
1\, -\, \left(1\, +\, \nu\, \frac{x}{\psi}\right)_+^{-1/\nu}
\qquad\mbox{for}\quad x>0, 
\label{eq:introGPD}
\end{equation}
where $y_+ = \max(y,0)$, and $\psi>0$ and $\nu\in\mathbb{R}$ are scale and shape parameters. The value of $\nu=0$, interpreted as the limit of \eqref{eq:introGPD} as $\nu\to 0$, gives the exponential distribution, whilst $\nu<0$ corresponds to a short-tailed distribution with finite upper end point $\mbox{$u-\psi\, /\, \nu$}$, and $\nu>0$ gives a power-law tail decay, heavier than that of a normal or gamma distribution. This family of distributions arises as the only possible non-degenerate limit for scaled excesses of a continuous random variable as the threshold tends to the upper end point of the distribution \citep{Pickands1975}. In practice, we assume that the limit has been sufficiently achieved above a large enough choice for $u$. Other than in hydrology, these models are applied in a range of areas, e.g., climatology \citep{Blanchet2011, Reich2014} and finance \citep{Chavez2014, Hilal2014}. We focus on modelling threshold exceedances, but we could have worked with an extremal mixture model \citep{Behrens2004,MacDonald2011} comprising separate distributions for observations below and above $u$ if the full distribution of $X$ was of interest.

When fitting extreme value models over a range of variables (e.g., the same physical variable measured at different sites, or different variables measured at the same time), it is natural to model the tails of the marginal distributions as GPD, with the parameters $(\psi, \nu)$ potentially changing over variables. A range of methods have been used to cluster the variables. Two broad approaches have emerged: methods that aim purely to find clusters of similarly distributed variables for purposes of interpretation; and methods that aim to pool information over similarly distributed variables to enhance inference efficiency. The first category of methods tend to evaluate extreme value theory summary statistics (e.g., the GPD shape parameter from each site) and apply widely used generic clustering techniques (e.g., k-means or k-medoids \citep{Kaufman2005}) to form clusters. Examples of such approaches include \citet{Rubio2018} for clustering different stocks based on extreme losses, and \citet{Bernard2013} and \citet{Bador2015} who explore different levels of pairwise dependence via clustering.

The latest approaches in the second category use hierarchical modelling. First versions of these methods go back to the pooling methods used by the Flood Studies Report (1975) \citep{Flood1975} which selected a hydrologically coherent region and assumed a common GPD shape parameter for all sites within the region. The methods evolved to also account for the dependence structure \citep{ColesTawn1990, ColesTawn1996}. These methods do not account for the uncertainty in the process of identifying the regions/clusters of similar variables. The first hierarchical method which accounted for the cluster uncertainty in the inference for extremes was by \citet{SmithGoodman2000}, later extended to a Bayesian mixture model by \citet{Bottolo2003}, who define a Bayesian framework to borrow information across multiple types of independent insurance claims in order to reduce uncertainty in the estimates of the parameters of the type-wise GPD. For spatial extreme value problems hierarchical clustering models for the extremes have been proposed by \citet{Carreau2017} and \citet{ReichShaby2018}, with both approaches having limitations. The former does not account for dependence over space. Although this feature is somewhat addressed by \citet{ReichShaby2018}, they model dependence over variables in the same cluster with a restrictive exchangeable parametric extreme value copula, which is likely to be too simplistic and as a consequence bias the marginal inference \citep{DupuisTawn2001}. Furthermore, the approach has a high computational cost and it fixes the number of clusters.  

The existing methods for spatial clustering of extremes focus solely on criteria based on the similarity of the marginal distributions, whereas others only consider dependence features. No methods look at both aspects yet, from a physical perspective, spatial dependence of a process is the key determinant of the marginal distributions being similar. Furthermore, the vast majority of approaches both ignore the clustering uncertainty (both numbers of clusters and the allocation of variables to clusters) and ignore the spatial dependence between variables within a cluster in their marginal inference. 

Our proposed approach is the first to address all of these features in a single Bayesian framework and it is applicable to both areal and geostatistical data. It learns about both the number of clusters and their spatial structure with respect to the site-wise distribution of the peaks-over threshold and spatial dependence in the extremes. Unlike \citet{ReichShaby2018} we do not attempt to estimate a full model for the spatial dependence structure, instead, similar to \citet{Bernard2013}, we account for dependence through a widely used pairwise measure of extremal dependence \citep{Coles1999}. We introduce a statistical model for the extremal dependence measure which incorporates both distance between sites and our belief that sites within the same cluster tend to exhibit a higher degree of dependence than sites in different clusters. Our approach to impose spatial structure on the parameters of the site-wise GPD is similar to \citet{Bottolo2003}, but the additional consideration of spatio-temporal dependence makes it a more general and harder problem.  Posterior samples are obtained using a reversible jump MCMC algorithm \citep{Green1995} which allows: the number of spatial clusters to vary, the analysis of the site-wise marginal tail behaviour, and the derivation of a point estimate for the cluster structure using Bayesian decision theory. 

The paper is organized as follows: Section~\ref{sec:Data} describes the motivating applications and introduces the data; Sections~\ref{sec:Method} and \ref{sec:Inference} detail the statistical modelling framework and the inference procedure; there is a simulation study in Section~\ref{sec:Example} and Section~\ref{sec:Application} presents a data analysis for the applications; and we conclude with a discussion in Section~\ref{sec:Discussion}.

\section{Motivating examples and data}
\label{sec:Data}

\subsection{Weather-related property insurance claims in Norway}
\label{sec:DataSN}

Insurance companies are interested in the distribution of the number of claims they are likely to receive over different areas as a consequence of rainfall and/or snow-melt. \citet{Scheel2013} showed that the upper tail of this distribution is exceptionally difficult to model for individual administrative areas (termed municipalities) in Norway. \citet{Rohrbeck2018} develop an extremal regression model that shows how the largest numbers of insurance claims per weather event in a municipality can be clearly linked to extreme precipitation in that municipality, and that from such a model, marginalizing over the precipitation gives a good model for the marginal distribution of the number of claims. 

While their approach achieves good results for three highly populated Norwegian municipalities, its complexity requires a high number of insurance claims in order to obtain reliable estimates. As this criterion is not satisfied by most Norwegian municipalities, there is a need to exploit spatial structure to borrow information across municipalities and instead model the total claims over clusters of municipalities. The question then is what precipitation value to use and how to model that. Although the probability of a claim given an extreme precipitation is likely to change very slowly over space, the distribution of extreme precipitation varies more rapidly due to geographical reasons. Therefore we need to identify clusters of municipalities which have both the same distribution of extreme precipitation and have similar actual values in each extreme event (i.e., strong extremal dependence), so that the average precipitation over the cluster can be used as a covariate for the aggregated claims across the municipalities within the cluster.

\begin{figure}
\includegraphics[height=7.5cm]{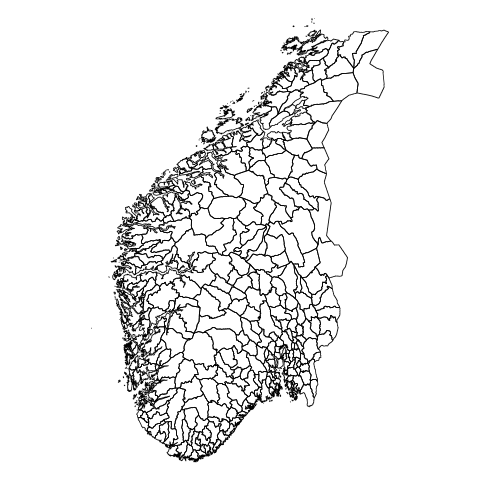}
\includegraphics[height=7.5cm]{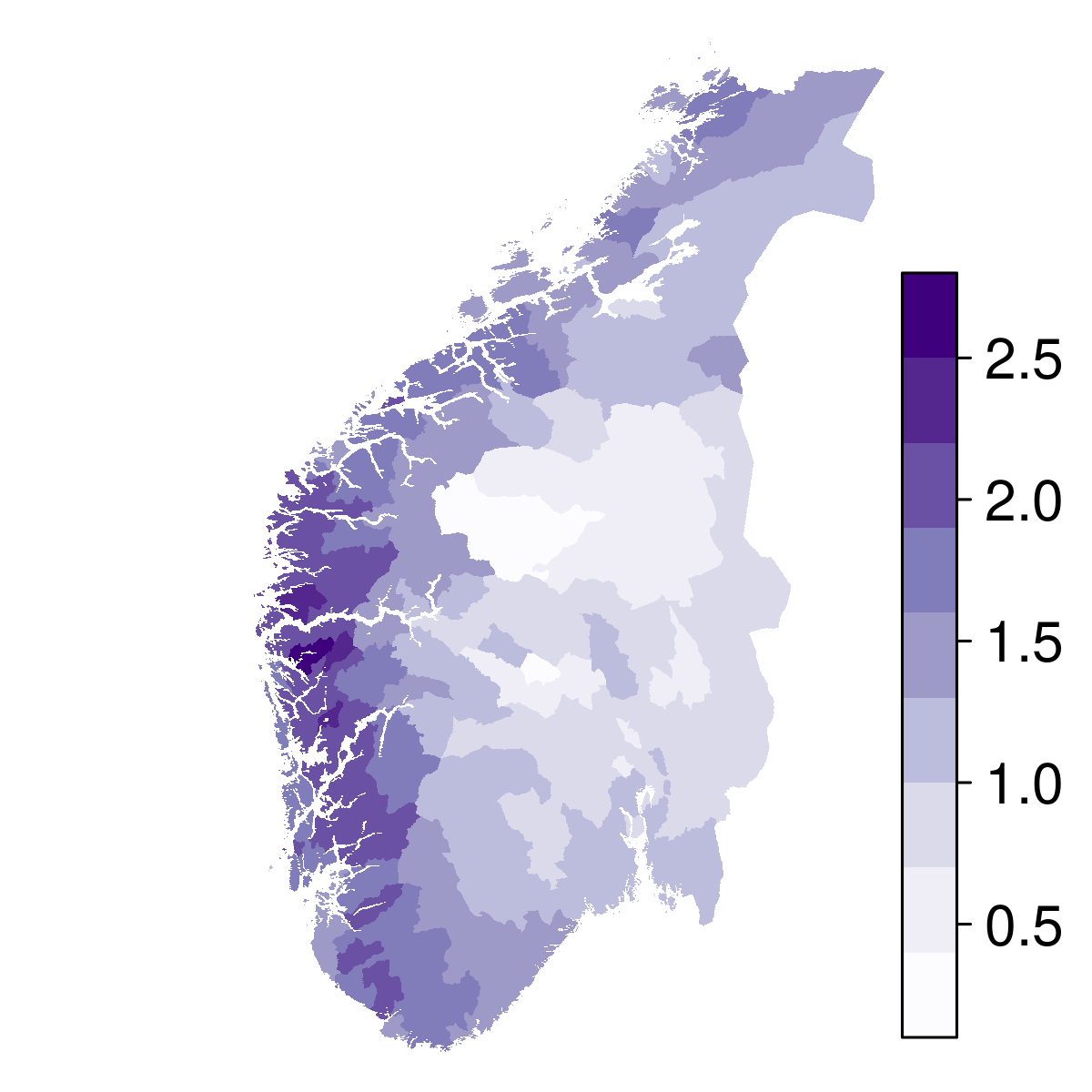}
\caption{Map of South Norway showing the boundaries of the municipalities (left) and the average daily amount of precipitation (mm) on log-scale between 1997 and 2006 (right).}
\label{fig:MapSN}
\end{figure}

We consider precipitation across the 343 municipalities in South Norway. These areal units, shown in  Figure~\ref{fig:MapSN} left panel, differ substantially in size, ranging from a few through to several hundred square kilometres. The data were produced by the Norwegian Meteorological Institute (\texttt{www.met.no}) and provide the daily amount of precipitation (in mm), including both rain and snowfall, between 1997 and 2006. Data for each municipality are obtained by a two-step process. First, point observations from more than 200 measurement stations across Norway are spatially interpolated to a regular grid of 1km$^2$. Then, for each municipality, the precipitation is obtained by a weighted averaging over the grid within the municipality's boundary, with weights proportional to the population density \citep{Haug2011}. The data exhibit a small number of missing values; 326 of the 343 municipalities have a full record and only six have more than 20 observations missing.

Figure~\ref{fig:MapSN} right panel shows substantial spatial variation in the average daily  precipitation across municipalities. The highest average values are recorded along the west coast and the averages decrease typically with increasing easterly coordinates and distance from the coast. All municipalities exhibit seasonality, with the largest average daily precipitation typically in September and October, but with the west coast having higher average precipitation levels in January to March, while these are the driest months, on average, for the municipalities in the south-east. 

To focus on the larger events, we explore the exceedances of a threshold, corresponding to the annual 90\% quantile for each municipality. For each municipality, the average number of events exceeding the threshold varies across the year, however, the excess values themselves are found not to exhibit seasonality. Similar results are found for all higher thresholds. Thus site-wise peaks over threshold can be considered as identically distributed over time but they do exhibit spatial variation.

\subsection{Flood risk analysis for the UK}
\label{sec:DataRF}

\begin{figure}
\begin{center}
\includegraphics[width=0.475\textwidth]{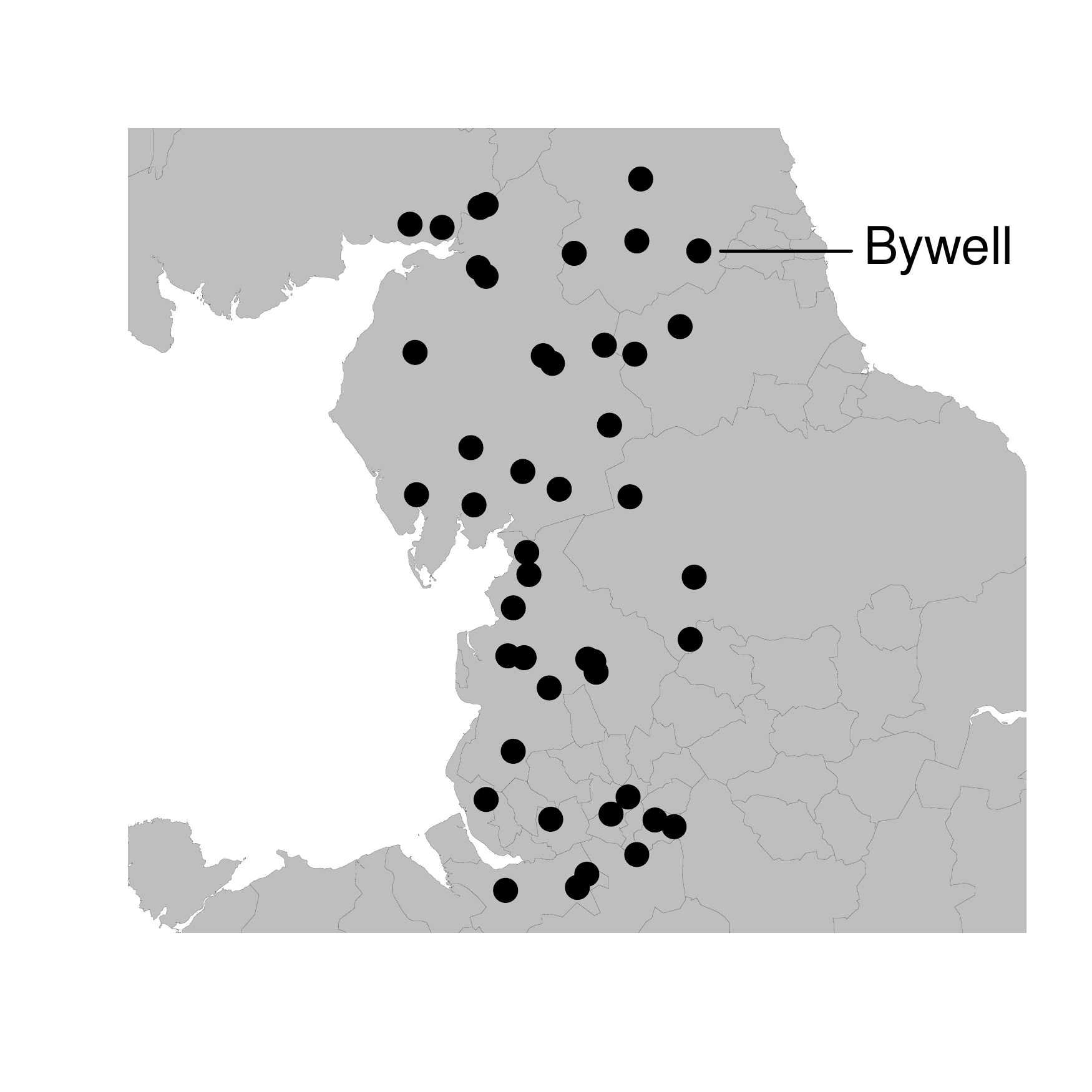}
\hspace{0.5cm}
\includegraphics[width=0.475\textwidth]{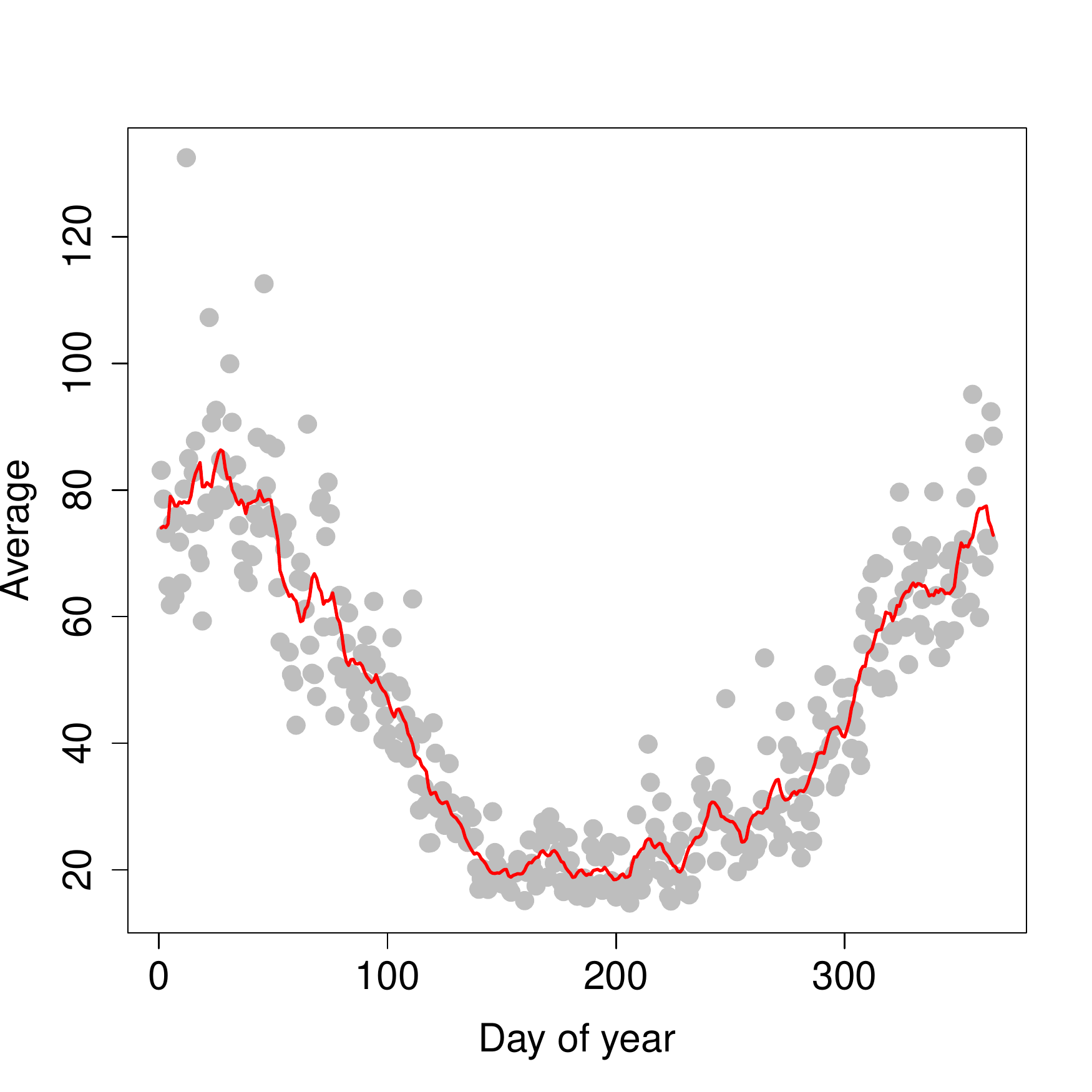}
\caption{Locations of the 45 river flow gauges considered in Section~\ref{sec:DataRF} (left) and the daily average river flow for the Bywell gauge on the River Tyne (right). The solid line in the right panel shows the two-week moving average over the year for Bywell.}
\label{fig:DataRFSites}
\end{center}
\end{figure}

Improving river flood risk analysis in the UK is of high priority given that it has experienced several severe and widespread flood events over the past years. For instance, the floods related to Storm Desmond, Storm Eva and Storm Frank in 2015/2016 caused an estimated economic damage of between \pounds1.3--1.9 billion \citep{UKES2018}. Practitioners are interested in the potential size of future flood events, as well as their spatial extent. Detecting groups of sites which have similar dynamics in terms of extreme river flow levels helps to address this question as we can combine them in the inference to improve statistical efficiency and hence produce estimates that give more reliable extrapolations to rarer events. However, extreme rainfall events in different seasons have different spatial characteristics due to frontal rainfall in summer and convective rainfall in winter; so clusterings may differ between seasons. 

Daily mean river flow levels (in m$^3/$s) for the years 1980 to 2011 for 45 gauges are obtained from the UK's National River Flow Archive (\texttt{nrfa.ceh.ac.uk}) and cover northern England and southern Scotland, with the majority being in North West England; see Figure~\ref{fig:DataRFSites} left panel. Observations for several years are missing for two gauges and other stations also exhibit some missing values. The data include the river flow levels for the floods in Cumbria in November 2009. Hydrological distances which account for catchment closeness are available; these provide a better emulation of the spatial dependence than the geographical distances \citep{Asadi2015}. 

Figure~\ref{fig:DataRFSites} right panel indicates strong seasonality for one of the gauges, a feature typical of all gauges in the region; the highest average river flow levels are observed for November through to March, while June through to August record the lowest averages. As such, a river flow level which is considered very high in summer may be rather standard in winter. Moreover, the data exhibit strong autocorrelation. Therefore, an extreme weather event may cause extreme river flow levels over consecutive days. The data values vary substantially across gauges; the site-wise daily average ranges from 0.2m$^3/$s up to 54.9m$^3/$s. All these aspects, that is seasonality, spatio-temporal dependence and difference in scale, are considered in our analysis in Section~\ref{sec:Application}. 

\section{Cluster model}
\label{sec:Method}

\subsection{Introduction}
\label{sec:MethodIntro}

Consider $K$ sites with spatial locations $\mathbf{s}_1,\ldots,\mathbf{s}_K\in \mathbb{R}^2$. For the Norwegian municipalities in Section~\ref{sec:DataSN}, $\mathbf{s}_k$~($k=1,\ldots,K$) refers to the centroid of the $k$-th areal unit (municipality). In a geostatistical setting, $\mathbf{s}_k$ is the point location of the $k$-th site; for instance, the latitude and longitude of the $k$-th gauge in Section~\ref{sec:DataRF}. Spatial proximity of any pair $(k,k')$ of sites is measured via a suitable metric which provides a distance $d_{k,k'}\geq 0$ based on the locations $\mathbf{s}_k$ and $\mathbf{s}_{k'}$. For each of the $K$ sites, we have data for the variable we wish to draw inference on the distribution of its largest values. 

The exploratory analysis in Section~\ref{sec:Data} shows that hydrological processes usually exhibit seasonality and spatio-temporal dependencies. As such, observations may neither be identically distributed nor independent. Firstly, we account for temporal dependence. We use declustering prior to any model fitting, with each of the $K$ time series being split into subperiods such that extremes occurring in different subperiods can be assumed to be independent \citep{FerroSegers2003}. We select the same set of subperiods for each site and only use the highest observation per subperiod and site in our analysis. Our methodology can be adapted to accommodate temporal lags between the same event at different sites but, in what follows, for notational simplicity, we present our method under the assumption that extreme events affecting multiple sites jointly have no time lag. 

Let $R_{k,1},\ldots,R_{k,T}$ denote the time series for site $k$~($k=1,\ldots,K$) after declustering, i.e., $R_{k,t}$ and $R_{k',t'}$ are assumed to be independent for any $t\neq t'$ and all $k$ and $k'$ including $k=k'$; the record length $T$ is taken to be equal across sites for simplicity. If $\mathbf{s}_k$ and $\mathbf{s}_{k'}$ are close together, corresponding to $d_{k,k'}$ being small, it is reasonable to assume that the marginal distributions for $R_{k,t}$ and $R_{k',t}$ should be the same, or very similar, therefore they should have similar GPD parameter values in~\eqref{eq:introGPD}. Spatial dependence leads to the same extreme event being present at different sites, again with the occurrence rate for such events likely to increase if $d_{k,k'}$ is small, i.e., if $R_{k,t}$ is large then the chances of $R_{k',t}$ being large increases if $\mathbf{s}_k$ and $\mathbf{s}_{k'}$ are close, relative to them being further apart. In Section~\ref{sec:ChiClusterModel} we introduce a pairwise extremal dependence measure, $\chi_{k,k'}$, between sites $k$ and $k'$ that captures this feature. 

Our approach is to group sites into clusters such that sites in the same cluster have both similar marginal distributions and the spatial dependence is greater between sites in the same cluster than between sites in different clusters. To represent the cluster structure, we introduce $K$ latent random variables $\mathbf{Z}=\left(Z_1,\ldots,Z_K\right)$. Let $J\in\left\{1,\ldots,K\right\}$ denote the number of clusters. Then, for each $k=1, \ldots ,K$,  $Z_k\in\left\{1,\ldots,J\right\}$ with $Z_k=j$ corresponding to the $k$-th site being allocated to the $j$-th cluster. Conditional on $\mathbf{Z}$, we propose separate models for the marginal distributions of $R_{k,t}$ and the dependence measures $\chi_{k,k'}$ in Sections~\ref{sec:GPDClusterModel} and \ref{sec:ChiClusterModel} respectively. We later combine these models in Section~\ref{sec:Inference}.

\subsection{Model for Marginal Clustering}
\label{sec:GPDClusterModel}

We model the site-wise tail behaviour using the GPD model~\eqref{eq:introGPD} for threshold exceedances. Our data in Section~\ref{sec:Data} indicates that $R_{k,1},\ldots,R_{k,T}$ are not identically distributed due to seasonality. There are a range of established approaches to handle this: to split the time series into shorter time periods for which the assumption of stationarity seems reasonable;  model the threshold and GPD parameters in terms of a set of temporal predictors \citep{Davison1990,Chavez2005}; to preprocess the data to remove non-stationary to the overall series, via the Box-Cox transformation \citep{Eastoe2009}. We will discuss the combination of these methods for our approach in more detail in Sections~\ref{sec:Application} and \ref{sec:Discussion}.  However, for simplicity, when presenting the methods we assume that $R_{k,1},\ldots,R_{k,T}$ are identically distributed. 

For site $k$, given a high enough threshold $u_k$, we model $R_{k,t}-u_k \mid(R_{k,t}>u_k)$ as following a GPD$(\psi_k,\nu_k)$. The selection of the threshold $u_k$ in \eqref{eq:introGPD} is highly important. If $u_k$ is too high, estimates for $\psi_k$ and $\nu_k$ are based on a small number of data points. Conversely, the distribution of $R_{k,t}-u_k\mid (R_{k,t}>u_k)$ may not be well approximated by a GPD if $u_k$ is too low. In many applications, $u_k$ is selected using graphical diagnostic tools with the mean residual life and threshold stability plots being the most common ones \citep{Coles2001}. More recent techniques are described by \citet{Wadsworth2016} and \citet{Northrop2017}. 

For clustering we group sites that have a very similar marginal tail behaviour. Therefore, the peaks-over threshold for all the sites within a cluster are modelled by a GPD with the same scale and shape parameters. The distribution of the peaks-over threshold for the $k$-th site ($k=1,\ldots,K$), conditional on $Z_k$, is thus given by
\begin{equation}
R_{k,t}-u_k\mid\left(Z_k=j\, ,\, R_{k,t}>u_k\right)~\sim~
\mbox{GPD}\left( \sigma_j\, ,\, \xi_j\right),
\label{eq:GPDCluster}
\end{equation}
where $\sigma_j>0$ and $\xi_j\in\mathbb{R}$~($j=1,\ldots,J$) denote the cluster-specific scale and shape parameters, therefore $(\psi_k,\nu_k)=(\sigma_j,\xi_j)$ if $Z_k=j$. We denote the parameters of the model given $\mathbf{Z}$ by $\bm\theta_\mathrm{M}^{(J)}=\left\{\bm\sigma^{(J)}, \bm\xi^{(J)}\right\}$, where  $\bm\sigma^{(J)}=(\sigma_1,\ldots,\sigma_J)$ and $\bm\xi^{(J)}=(\xi_1,\ldots,\xi_J)$. 

\subsection{Model for Spatial Dependence Clustering}
\label{sec:ChiClusterModel}

There are a number of ways for modelling dependence in multivariate and spatial extremes. Multivariate approaches have included fitting parametric models for multivariate extreme value copula \citep{Tawn1988} and various threshold methods \citep{Ledford1997,Ferreira2014, Rootzen2018}. In a spatial context, max-stable processes are the most widely used \citep{Davison2012,ReichShaby2012, Asadi2015}. Such processes arise as the limit of component-wise maxima of suitably normalized replications of a spatial process \citep{Haan1984} and multiple parametric representations have been proposed \citep{Davison2012}. However, they have a number of inference and model limitations. Inference issues are mostly overcome by instead using Pareto processes \citep{Ferreira2014, Dombry2015}; however these models give either a strong form of extremal dependence (corresponding to $\chi_{k,k'}>0$ in \eqref{eq:Chi}) across all sites or give independence. Alternative methods that allow for spatial dependence that weakens with extremal level also exist \citep{Wadsworth2012b, Wadsworth2018}. 

None of these methods has both the sufficient flexibility and ease of implementation to be applied reliably to problems of the dimension of those in Section~\ref{sec:Data}. Therefore, instead of attempting to model the full joint distribution over the extreme events at the sites, we model the core summary statistics for pairwise extremal dependence. The most widely used extremal dependence measure is the coefficient of asymptotic dependence $\chi$ \citep{Coles1999}. Formally, for the random variables $R_k$ and $R_{k'}$~($k,k'=1,\ldots,K$), $\chi_{k,k'} = \lim_{u\to1}\chi_{k,k'}(u)$ where
\begin{equation}
\chi_{k,k'}(u)~=~\mathbb{P}\left[\, F_{k}(R_k) > u\mid F_{k'}(R_{k'}) > u\, \right],\qquad u\in[0,1],
\label{eq:Chi}
\end{equation}
with $F_{k}(\cdot)$ and $F_{k'}(\cdot)$ denoting the cumulative distribution functions of $R_k$ and $R_{k'}$, respectively. Thus $\chi_{k,k'}\in[0,1]$ gives the limit probability of site $k$ observing an extreme event conditional on site $k'$ recording one. There are strong parallels between $\chi_{k,k'}$ and the extremogram \citep{Davis2009} at distance $d_{k,k'}$, with the key difference being that inference for the extremogram pools together all the data from pairs of sites with the same separation under the assumption of stationarity. When $\chi_{k,k'}>0$, $R_k$ and $R_{k'}$ are termed asymptotically dependent, with increasing values corresponding to stronger extremal dependence, whilst we say that $R_k$ and $R_{k'}$ are asymptotically independent if $\chi_{k,k'}=0$. Asymptotic dependence can, for instance, arise when the conditions for bivariate regular variation hold \citep{Resnick2013}, while random variables with a Gaussian copula are asymptotically independent unless they are perfectly dependent. 

Before attempting to specify the model for the distribution of $\chi_{k,k'}$, conditional on $\mathbf{Z}$, consider what properties the expected value of this conditional distribution should possess. Since sites $(k,k')$ within the same cluster are expected to have a higher probability of joint extreme events, $\chi_{k,k'}\mid\mathbf{Z}$ should be larger for $Z_k = Z_{k'}$ than when $Z_k \neq Z_{k'}$. We express this assumption via the constraint
\begin{equation}
\mathbb{E}\left(\chi_{k,k'} \mid Z_k = Z_{k'}\right)~\geq~ \mathbb{E}\left(\chi_{k,k'} \mid Z_k \neq Z_{k'}\right).
\label{eq:Constraint}
\end{equation}
Furthermore, the pairwise extremal dependence between the sites $k$ and $k'$, irrespective of whether the sites are in the same cluster or not, should decrease with increasing distance $d_{k,k'}$.  We assume that the expectation of $\chi_{k,k'}$ decays exponentially with $d_{k,k'}$, but with a varying rate across each cluster and with a common, but faster, decay rate between clusters. These properties are reflected in the formulation
\begin{equation}
\mathbb{E}\left(\chi_{k,k'} \mid \mathbf{Z}\right) ~=~
\begin{cases}
\exp\left(-\gamma_j ~ d_{k,k'} \right) & \quad\mbox{if}\quad Z_k=Z_{k'}=j,\\
\exp\left(-\gamma_0 ~ d_{k,k'} \right) & \quad\mbox{if}\quad Z_k\neq Z_{k'},\\
\end{cases}
\label{eq:ChiCluster}
\end{equation}
where $\gamma_j>0$~($j=1,\ldots,J$) is a cluster-specific parameter and $\gamma_0>\max(\gamma_1,\ldots,\gamma_J)\geq 0$. This approach is consistent with $\chi_{k,k}=1$~($k=1,\ldots,K)$ and constraint \eqref{eq:Constraint}, and allows for non-stationarity of the process as $\gamma_i$ and $\gamma_j$, $i\neq j\geq 1$, can differ.  We ensure that $\gamma_0>\max(\gamma_1,\ldots,\gamma_J)$ by defining parameters $\epsilon_1,\ldots,\epsilon_J$ such that
\begin{equation}
\log\left(\gamma_j\right) ~=~ \log(\gamma_0)\, -\, \epsilon_j, \qquad\epsilon_j\geq0, \qquad (j=1,\ldots,J).
\label{eq:LikeGamma}
\end{equation}

Now consider the distribution of $\chi_{k,k'}\mid\mathbf{Z}$. As $\chi_{k,k'}\in[0,1]$ may differ between two pairs of sites in the same cluster with the same $d_{k,k'}$, due to factors such as topology, we choose a beta distribution model with
\begin{equation}
\chi_{k,k'} \mid \mathbf{Z} ~\sim~
\begin{cases}
\mbox{Beta}\left(\frac{\beta \exp\left(-\gamma_j~d_{k,k'} \right)}{1 - \exp\left(-\gamma_j ~ d_{k,k'} \right)}~,~\beta\right) & \quad\mbox{if}\quad Z_k=Z_{k'}=j,\\
\vspace{-0.5cm}\\
\mbox{Beta}\left(\frac{\beta \exp\left(-\gamma_0~d_{k,k'} \right)}{1 - \exp\left(-\gamma_0 ~ d_{k,k'} \right)}~,~\beta\right) & \quad\mbox{if}\quad Z_k\neq Z_{k'},\\
\end{cases}
\label{eq:ChiClusterBeta}
\end{equation}
which has expectation given by \eqref{eq:ChiCluster} and where $\beta>0$ controls the variance of $\chi_{k,k'}$; higher values of $\beta$ correspond to $\chi_{k,k'}$ being less variable. Consequently, the spatial variation of the set of coefficients of asymptotic dependence, conditional on $\mathbf{Z}$, is defined via the $J+2$ parameters $\gamma_0$, $\bm\epsilon^{(J)} = (\epsilon_1,\ldots,\epsilon_J)$ and $\beta$, and we denote the dependence parameters given $\mathbf{Z}$ by $\bm\theta_\mathrm{D}^{(J)}=(\gamma_0,\bm\epsilon^{(J)},\beta)$.

The distribution \eqref{eq:ChiClusterBeta} could be applied to model $\chi_{k,k'}$ for all pairs of sites $(k,k')$. However, this may not be optimal. For $Z_{k}\neq Z_{k'}$, $\chi_{k,k'}$ may vary strongly, depending on whether sites $k$ and $k'$ belong to adjacent clusters or whether there are multiple clusters along the path between the two sites. Such differences can probably not be captured by the single parameter $\gamma_0$. We thus consider $\chi_{k,k'}$ for adjacent pairs of sites only. In case of the sites being point locations, we first derive the Voronoi partition of the study area and then define sites as being adjacent if their corresponding Voronoi cells are adjacent.

\section{Bayesian Inference} 
\label{sec:Inference}

\subsection{Introduction}
We use Bayesian inference for the number of clusters, $J$, the latent variables  $\mathbf{Z}$, and the marginal and dependence structure parameters $\bm\theta_\mathrm{M}^{(J)}$ and $\bm\theta_\mathrm{D}^{(J)}$ using the declustered data $\mathcal{D}= \left\{\, \left(r_{k,1},\ldots,r_{k,T}\right)\, :\, k=1,\ldots,K\, \right\}$. The posterior, and the algorithm to sample from it, are developed in this section. The derivation of the marginal and dependence structure likelihood contributions $L_{\mathrm{M}}$ and $L_{\mathrm{D}}$, given $\mathbf{Z}$ and $\mathcal{D}$, are provided in Section~\ref{sec:ClusterInference}.  Critically, the data that are used to analyze the marginal and dependence structures are different, in that for marginal distributions we use all marginal exceedances of a threshold at all sites, whereas for the dependence model, only ranks of the variables are used. These two forms of the data are only weakly dependent. Furthermore, \citet{Genest1995} and \citet{Genest2009} have shown that inference for dependence parameters is largely unaffected by marginal parameter estimation. Therefore we use the following approximation to the joint likelihood, the independent likelihood
\begin{equation}
L\left(\bm\theta_\mathrm{M}^{(J)}, \bm\theta_\mathrm{D}^{(J)}\mid\mathcal{D},\mathbf{Z}\right)~=~
L_{\mathrm{M}}\left(\bm\theta_\mathrm{M}^{(J)} \mid\mathcal{D},\mathbf{Z}\right) 
\, \times\, L_{\mathrm{D}}\left(\bm\theta_\mathrm{D}^{(J)} \mid\mathcal{D},\mathbf{Z}\right)\hspace{-0.1cm}.
\label{eq:LikelihoodAreal}
\end{equation}
Section~\ref{sec:GeoData} presents our priors, including a spatial prior for $\mathbf{Z}$ given $J$ and a prior for $J$. Section~\ref{sec:Estimation} details our algorithm to sample from the posterior distribution, and Section~\ref{sec:Pointestimate} outlines the analysis of the posterior samples and the estimation of the underlying cluster structure.

\subsection{Likelihood Components}
\label{sec:ClusterInference}

\subsubsection{Marginal component}
\label{sec:GPDClusterInference}

If the peaks-over threshold data were independent over all sites the likelihood function for 
$\bm\theta_\mathrm{M}$, conditional on $\mathbf{Z}$, the data $\mathcal{D}$ and thresholds $\mathbf{u}=(u_1,\ldots,u_K$) would be
\begin{equation}
L_{\mathrm{M}}^{\mathrm{ind}}\left(\bm\theta_\mathrm{M}^{(J)}\mid\mathcal{D},\mathbf{Z}\right)~=~
\prod_{k=1}^K \prod_{\left\{t~:~r_{k,t}>u_k\right\}}
~\frac{1}{\sigma_{Z_k}}
\left(1+\xi_{Z_k}\, \frac{r_{k,t}-u_k}{\sigma_{Z_k}}\right)_+^{-1/\xi_{Z_k}-1}.
\label{eq:Likemisspecified}
\end{equation} 
However, spatial independence is not a valid assumption for our motivating problems since severe weather events usually affect a number of sites (municipalities/gauges). Therefore, the likelihood function in expression~\eqref{eq:Likemisspecified} corresponds to a misspecified model. Under suitable regularity conditions, \citet{Kent1982} shows that a general theory for the asymptotic distribution of the maximum likelihood estimate $\hat{\bm\theta}_\mathrm{M}^{(J)}$ based on the misspecified likelihood~\eqref{eq:Likemisspecified} is
\begin{equation}
\sqrt{T}\left(\hat{\bm\theta}_\mathrm{M}^{(J)} - \bm\theta_\mathrm{M}^{(J)} \right)~\sim~ 
\mbox{Normal}\left(\, \mathbf{0}\, ,\, \Sigma=H\left(\bm\theta_\mathrm{M}^{(J)}\right)^{-1} V\left(\bm\theta_\mathrm{M}^{(J)}\right) H\left(\bm\theta_\mathrm{M}^{(J)}\right)^{-1}\, \right),
\label{eq:Kent1982}
\end{equation}
where $\bm\theta_\mathrm{M}^{(J)}$ are the true model parameters, 
$H\left(\bm\theta_\mathrm{M}^{(J)}\right)=-\mathbb{E}\left[\nabla^2 \log L_\mathrm{M}^{\mathrm{ind}}\left(\bm\theta_\mathrm{M}^{(J)}\mid\mathcal{D},\mathbf{Z}\right)\right]$ denotes the Fisher information, $V\left(\bm\theta_\mathrm{M}^{(J)}\right) = 
\mbox{Cov}\left[\nabla\log L_\mathrm{M}^{\mathrm{ind}}\left(\bm\theta_\mathrm{M}v\mid\mathcal{D},\mathbf{Z}\right)\right]$ and $\nabla^i$ refers to the $i$-th derivative. The limiting variance in~\eqref{eq:Kent1982} is different from the classic Fisher information if the model was misspecified, but that if we have no spatial dependence, $H\left(\bm\theta_\mathrm{M}^{(J)}\right) = V\left(\bm\theta_\mathrm{M}^{(J)}\right)$ and then the classic asymptotic result is obtained. Inference under the misspecified model, when there is spatial dependence would underestimate the variance of the estimator $\hat{\bm\theta}_\mathrm{M}^{(J)}$. With respect to our Bayesian framework, this would correspond to credible intervals of the parameters being too narrow.
 
We follow an approach of \citet{Ribatet2012}, who propose an adjustment of the curvature of the likelihood \eqref{eq:Likemisspecified} around its mode using the asymptotic behaviour in expression~\eqref{eq:Kent1982}. The adjusted likelihood is 
\begin{equation}
L_\mathrm{M}^\mathrm{adj}\left(\bm\theta_\mathrm{M}^{(J)}\mid\mathcal{D}, \mathbf{Z}\right) ~=~ 
L_\mathrm{M}^{\mathrm{ind}}\left(\hat{\bm\theta}_\mathrm{M}^{(J)} + B\left(\bm\theta^{(J)}_\mathrm{M}- \hat{\bm\theta}_\mathrm{M}^{(J)}\right) \mid\mathcal{D}, \mathbf{Z}\right), 
\label{eq:LikeAdj}
\end{equation}
where the $2J\times 2J$ matrix $B$ depends on $\mathbf{Z}$ and is 
\begin{equation}
B ~=~ \left\{\left[H\left(\bm\theta_\mathrm{M}^{(J)}\right)\right]^{1/2}\right\}^{-1} \left[\Sigma^{-1}\right]^{1/2}
\label{eq:Bj}
\end{equation}
with $\left[\cdot\right]^{1/2}$ denoting the matrix square root. To compute $B$, the evaluation of $H$ and $V$ matrices uses the observed information matrix and an estimate for the covariance matrix of the score function respectively, both evaluated with $\bm\theta_\mathrm{M}^{(J)} = \hat{\bm\theta}_\mathrm{M}^{(J)}$. Note, $B$ is a block diagonal matrix consisting of $J$ lots of $2\times2$ blocks, each block corresponds to one cluster (i.e.\ for the $j$-th block this corresponds to the terms for $\sigma_j$ and $\xi_j$); this feature enables efficient computation of $B$.  

Using this adjusted likelihood has a number of key properties: not changing the maximum likelihood estimate relative to using likelihood $L_\mathrm{M}^{\mathrm{ind}}$, suitably inflating the variances to be consistent with \eqref{eq:Kent1982} when there is spatial dependence, and leaving the inference unchanged from using likelihood $L_{\mathrm{M}}^{\mathrm{ind}}$ in the absence of spatial dependence. We therefore take our likelihood contribution for the marginal parameters to be $L_\mathrm{M}\left(\bm\theta_{\mathrm{M}}^{(J)}\mid\mathcal{D},\mathbf{Z}\right)=L_\mathrm{M}^{\mathrm{adj}}\left(\bm\theta_\mathrm{M}^{(J)}\mid\mathcal{D}, \mathbf{Z}\right)$.

\subsubsection{Dependence component}
\label{sec:ChiClusterInference}

With $\chi_{k,k'}(u)$ defined as in expression~\eqref{eq:Chi}, we assume that there exists a value $0\le \tilde{u}<1$ such that $\chi_{k,k'}(u) = \chi_{k,k'}$, i.e., it is equal to its limit form, for all $\tilde{u}<u<1$ and for all pairs of sites $(k,k')$. Techniques for the selection of $\tilde{u}$, for a pair $(k,k')$, are available \citep{deHaan1998,Wan2018}. Our data for estimating $\chi_{k,k'}$ are derived from exceedances of the 100$\tilde{u}$\% quantiles at sites $k$ and $k'$, as determined by their respective empirical distribution estimates $\hat{F}_k$ and $\hat{F}_{k'}$. First let $\mathcal{Q}_{k,k'} = \left\{t\, :\, \hat{F}_{k'}(r_{k',t})>\tilde{u}\right\}$, be the set of times when there is both an exceedance of the quantile threshold at site $k'$ and the data for site $k$ are available, and let $Q_{k,k'} = \lvert\mathcal{Q}_{k,k'}\rvert$ denote the cardinality of this set.  Further let 
\begin{equation}
P_{k,k'}~=~\#\left\{t\in \mathcal{Q}_{k,k'} \, :\, \hat{F}_{k}(r_{k,t})>\tilde{u}\right\},\qquad k,k'=1,\ldots,K.
\end{equation}
Then it follows that $P_{k,k'}$ is distributed as
\begin{equation}
P_{k,k'}\mid\chi_{k,k'} ~\sim~ \mbox{Binomial}\left(\, Q_{k,k'}\, ,\, \chi_{k,k'}\, \right)\hspace{-0.1cm}.
\label{eq:Pkk}
\end{equation}
In the absence of knowledge of our model in Section~\ref{sec:ChiClusterModel} for $\chi_{k,k'}$ over sites, the estimate for $\chi_{k,k'}$ is given by
\begin{equation}
\hat{\chi}_{k,k'} ~ = ~ \frac{P_{k,k'}}{Q_{k,k'}}, 
\label{eq:ChiEmpirical}
\end{equation}
which, if there were no missing data, is simply the proportion of the exceedances of the 100$\tilde{u}$\% for site $k'$ that also exceed this quantile threshold for site $k$. So estimator~\eqref{eq:ChiEmpirical} is an extension of the standard estimator for $\chi_{k,k'}$. Note that due to the missing data this estimator has the property that
$\hat{\chi}_{k,k'} $ does not necessarily equal $\hat{\chi}_{k',k} $ even though $\chi_{k,k'} =\chi_{k',k}$.

By combining \eqref{eq:Pkk} and the information about $\chi_{k,k'}$, from our cluster model~\eqref{eq:ChiClusterBeta}, and then integrating over $\chi_{k,k'}$, we obtain that $P_{k,k'} \mid (\mathcal{D}, \mathbf{Z})$ follows a beta-binomial distribution of the form
\begin{equation}
P_{k,k'} \mid \mathcal{D},\mathbf{Z} ~\sim~
\begin{cases}
\mbox{Beta-binomial}
\left(Q_{k,k'}~,~\frac{\beta}{\exp\left(\gamma_j d_{k,k'}\right)-1}~,~\beta\right)
& \quad\mbox{if}\quad Z_k=Z_{k'}=j,\\
\vspace{-0.5cm}\\
\mbox{Beta-binomial}
\left(Q_{k,k'}~,~\frac{\beta}{\exp\left(\gamma_0 d_{k,k'}\right)-1}~,~\beta\right)
& \quad\mbox{if}\quad Z_k\neq Z_{k'}.\\
\end{cases}
\label{eq:PkkCluster}
\end{equation}

Following the discussion in Section~\ref{sec:ChiClusterModel} this model is assumed to hold only for
pairs of sites $(k,k')$ which are adjacent. We let $k\sim k'$ denote the distinct pairs of sites $(k,k')$ which are adjacent. Denote the density function of the beta-binomial distribution in \eqref{eq:PkkCluster} by $g$, then for a pair of adjacent sites $(k,k')$ the likelihood contribution to $L_{\mathrm{D}}\left(\bm\theta_\mathrm{D}^{(J)}\mid\mathcal{D},\mathbf{Z}\right)$ is
\[
L_{\mathrm{D}}^{k,k'}=
\left[g\left(\, P_{k,k'} \mid Q_{k,k'}, \mathbf{Z},\bm\theta_\mathrm{D}^{(J)}\, \right)\, \times\,
g\left(\, P_{k',k} \mid Q_{k',k}, \mathbf{Z},\bm\theta_\mathrm{D}^{(J)}\, \right)\right]^{0.5},
\]
as critically we have two estimates for $\chi_{k,k'}$ which contain almost exactly the same information, each of equal value, and so we weight both observations by $0.5$.

Under an assumption of independence of distinct pairs, the likelihood function for the spatial dependence model is then  
\begin{equation}
L_{\mathrm{D}}\left(\bm\theta_\mathrm{D}^{(J)}\mid \mathcal{D}, \mathbf{Z} \right)
~=~ \prod_{k\sim k'}\, L_{\mathrm{D}}^{k,k'}.
\label{eq:Ldep}
\end{equation}
This likelihood is misspecified since $\left\{P_{k,k'}\,:\,k\neq k'\right\}$ are not independently distributed; for instance, when there is no missing data, if $P_{k,k'}$ and $P_{k,k''}$ are very large, then $P_{k',k''}$ cannot be small. We could use the \citet{Ribatet2012} adjustment again but we opted against it due to the following reasons. Firstly, in our data examples, the values for $P_{k,k'}$ are not so close to $Q_{k,k'}$ as that the effect is very strong. Secondly, we are not directly interested in the marginal posterior distributions of $\bm\theta_\mathrm{D}^{(J)}$ and as for our posterior, given $\mathbf{Z}$ and $\mathcal{D}$, $\bm\theta_\mathrm{M}^{(J)}$ and $\bm\theta_\mathrm{D}^{(J)}$ are independent. Finally, the observed information matrix which we would require for the curvature adjustment is dense due to the parameter $\beta$ being present in all terms of the likelihood function and of dimension $(J+2) \times (J+2)$. Therefore, we would have to invert a potentially high-dimensional matrix, which leads to a substantial computational cost, in particular, since we have to compute this matrix many times. 
 
\subsection{Priors} 
\label{sec:GeoData}

To perform Bayesian inference, we specify priors for the number of clusters $J$, the cluster labels $\mathbf{Z}\mid J$, and the model parameters $\bm\theta_\mathrm{M}^{(J)}$ and $\bm\theta_\mathrm{D}^{(J)}$. Since $J\geq1$, we define $J-1\sim\mbox{Poisson}(\kappa)$ and set a weakly informative Gamma prior for $\kappa$, $\kappa\sim \mbox{Gamma}(1,0.001)$. 

We wish to impose that clusters are contiguous; otherwise, two sites which are far apart may be grouped together, despite the probability of them jointly facing an extreme event being close to zero. Our prior is similar to \citet{Held2000} and only gives positive mass to contiguous clusters. The idea is to represent the spatial structure of the $J$ clusters via a set of centres $\mathbf{C}^{(J)} = (C_1,\ldots,C_J)\in \left\{1,\ldots,K\right\}$, $C_i\neq C_j$ if $i\neq j$; $C_i=j$ corresponds to $\mathbf{s}_j$ being the centre of the $i$-th cluster. Each site $k$ is assigned to the closest cluster centre in terms of $d_{k,k'}$~($k,k'=1,\ldots,K$), i.e., we take 
\begin{equation}
Z_k \mid \mathbf{C}^{(J)} = \argmin_{j=1,\ldots,J}~d_{k,C_j}.
\label{eq:Clustering}
\end{equation}
To ensure that $\mathbf{Z}\mid\mathbf{C}^{(J)}$ is well-defined, the site $k$ is allocated to the cluster with lowest index if multiple cluster centres have minimum distance to the site. Relationship \eqref{eq:Clustering} implies that we can assign a prior for $\mathbf{Z}\mid J$ via one for $\mathbf{C}^{(J)}$. We choose a uniform prior with
\[
\mathbb{P}\left(\mathbf{C}^{(J)} \mid J\right) = \frac{(K-J)!}{K!}.  
\]

We conclude by assigning priors for the parameters $\bm\theta_\mathrm{M}^{(J)}$ and $\bm\theta_\mathrm{D}^{(J)}$. For the GPD parameters in \eqref{eq:GPDCluster}, a log-normal prior is defined for the scale parameter, $\sigma_j\sim\mbox{Lognormal}(\mu^{\sigma}, \theta^{\sigma})$, while a normal prior is set for the shape parameter, $\xi_j\sim\mbox{Normal}(\mu^{\xi}, \theta^{\xi})$~($j=1,\ldots,J$). The priors for the parameters describing the spatial dependence are defined as exponentially distributed. Specifically, $\epsilon_j \sim \mbox{Exponential}\left(\theta^{\epsilon}\right),~\gamma_0\sim\mbox{Exponential}(0.001)$ and  $\beta\sim\mbox{Exponential}(0.01)$. The exponential prior for $\epsilon_j$ represents a prior preference to small spatial differences in the extremal dependence. To complete the model setup, we specify independent conjugate priors for the hyperparameters: $\mu^{\sigma}\sim\mbox{Normal}(0,1)$,   
$\mu^{\xi}\sim\mbox{Normal}(0,0.2)$, $\theta^{\sigma},\theta^{\xi}\sim\mbox{Inverse-Gamma}(1,0.1)$ and $\theta^{\epsilon}\sim\mbox{Gamma}(5,2)$.

\subsection{Implementation}
\label{sec:Estimation}

We wish to sample from the posterior distribution defined by the likelihood function \eqref{eq:LikelihoodAreal} and the prior distributions in Sections~\ref{sec:GeoData}. Since the dimension of the parameter space changes with the number $J$ of clusters, we use a reversible jump MCMC algorithm (RJMCMC) \citep{Green1995}. Given a current sample with $J$ clusters, we propose one of the following seven moves:
\begin{description}
\item[Birth:] Introduce a new cluster centre $C^*$ with parameters $\epsilon^*$, $\sigma^*$ and $\xi^*$.
\item[Death:] Remove one of the existing cluster centres $C_1,\ldots,C_J$. Sites previously allocated to the removed cluster are assigned to the other $J-1$ clusters according to \eqref{eq:Clustering}.
\item[Shift:] Move one of the cluster centres to a nearby site which is not a cluster centre and update the cluster labels $\mathbf{Z}\mid J$ according to \eqref{eq:Clustering}.
\item[Sigma:] Update the scale parameters $\bm\sigma^{(J)}$ of the GPD in \eqref{eq:GPDCluster}.
\item[Xi:] Update the shape parameters $\bm\xi^{(J)}$ of the GPD in \eqref{eq:GPDCluster}.
\item[Chi:] Update $\bm\epsilon^{(J)}$, $\gamma_0$ and $\beta$ in \eqref{eq:PkkCluster}.
\item[Hyper:] Update the hyperparameters $\kappa$, $(\mu^{\sigma}, \mu^{\xi})$ and $(\theta^{\sigma}, \theta^{\xi}, \theta^{\epsilon})$.
\end{description}
The birth and death move are comparable to the split and merge moves defined in other Bayesian clustering approaches, see \citet{Bottolo2003} for instance, but they potentially affect more than one cluster. For the examples in Sections~\ref{sec:Example} and \ref{sec:Application}, birth, death and shift are proposed with probability 0.2 each while the remaining four moves are each proposed with probability 0.1.

We briefly describe some features of our implementation and more details are provided in Appendix \ref{sec:RJMCMC}. For a birth move, the new cluster centre $C^*$ is uniformly sampled from one of the $K-J$ sites which are not currently cluster centres. In addition to $C^*$, the index at which to insert $C^*$ in the vector $\mathbf{C}^J$ is sampled with equal probability. The proposal distributions for the cluster parameters $\epsilon^{*}$, $\sigma^{*}$ and $\xi^{*}$ are close to the priors in Section~\ref{sec:GeoData} but some spatial information is incorporated; in particular, the mean of the proposal distribution is set to a average of the current parameter values of the sites allocated to the new cluster, while the variance of the proposal distribution is set to the prior variance. A death move ensures reversibility and one of the existing $J$ cluster centres is removed with equal probability. If $J=1$, the death move is rejected immediately. A shift move reallocates an existing cluster centre to one of the adjacent sites which are not currently cluster centres; this changes neither the cluster parameters nor the indexing of the cluster centres. Note, the matrix $B$ in expression \eqref{eq:LikeAdj} has to be updated in the case of a birth, death or shift move; due to $B$ being a block diagonal matrix, we only have to update the $2\times2$ blocks of the clusters which are affected by the proposal. The model parameters in $\bm\theta_{\mathrm{M}}^{(J)}$ and $\bm\theta_{\mathrm{D}}^{(J)}$ are updated via a Metropolis-within-Gibbs algorithm, while we sample from the corresponding full conditional distributions when updating the hyperparameters. 

\subsection{Analysis of the posterior samples}
\label{sec:Pointestimate}

Interest lies in the marginal distributions of the peaks-over threshold and the underlying spatial cluster structure. Suppose that we generated $N$ posterior samples as described in Section~\ref{sec:Estimation} and let $Z_k^{(i)}$ be the cluster label for site $k$ in the $i$-th sample~($i=1,\ldots,N$). To obtain a point estimate for the cluster structure, we first derive the posterior probability of sites $k$ and $k'$~($k,k'=1,\ldots,K$) being in the same cluster. This gives $\mathbf{S}\in[0,1]^{K\times K}$ with $(k,k')$ entry
\begin{equation}
S_{k,k'} ~=~ \frac{1}{N}\sum_{i=1}^N \mathbbm{1}\left\{Z^{(i)}_{k} = Z^{(i)}_{k'}\right\}.
\label{eq:SimilarityMatrix}
\end{equation}
We estimate the cluster structure using the Bayesian decision theoretical approach by \cite{Wade2018}, provided in the \texttt{mcclust.ext} \texttt{R} package \citep{Wade2015}. They derive a point estimate for the cluster structure based on the variation of information criterion \citep{Meila2007} which was originally introduced for comparing clusterings.

The marginal distributions can be estimated by Monte Carlo integration. For instance, the posterior mean estimate for $\psi_k$ is
\begin{equation}
\hat{\psi}_k ~=~ \frac{1}{N}\sum_{i=1}^N \sigma_{Z_k^{(i)}}^{(i)},
\label{eq:MCI}
\end{equation}
where $\sigma_{Z_k^{(i)}}^{(i)}$ is the cluster-specific scale parameter for cluster $Z_k^{(i)}$. Similarly, we obtain credible intervals for the GPD scale and shape parameters. We refer to this approach as site-wise Monte Carlo integration (SWMC); this site-wise estimation is usually done in Bayesian cluster analysis. 

For illustration and comparison purposes, we consider a second approach to estimate the marginal distributions based on the posterior samples; we later apply it in Section~\ref{sec:Application}. The estimate first imposes the spatial clusters and then estimates the GPD cluster parameters using the likelihood $L_{\mathrm{M}}^{\mathrm{adj}}(\cdot)$ in \eqref{eq:LikeAdj}. We fix the clusters as the point estimate above and assume sites within a cluster to have the same scale and shape parameter (as it is done in many regional hazard models). Let $J^*$ denote the estimated number of clusters and let $\mathcal{C}_1,\ldots,\mathcal{C}_{J^*}$ be such that $\mathcal{C}_j$ contains the sites allocated to the $j$-th cluster in the point estimate. We then consider the posterior $(\bm\sigma^{(J^*)},\bm\xi^{(J^*)}\mid\mathbf{Z}^*,\mathcal{D})$, where $\mathbf{Z}^*$ represents $\mathcal{C}_1,\ldots,\mathcal{C}_{J^*}$. Posterior samples are generated using our implementation in Section 4.4 with birth, death and shift being proposed with zero probability. The cluster-specific parameters $(\sigma_j,\xi_j)$ for cluster $j$~($j=1,\ldots,J^*$) are then estimated by Monte Carlo integration with
\[
\hat{\sigma}_j~=~\frac{1}{N}\sum_{i=1}^N \sigma_{j}^{(i)},
\]
where $N$ is the sampled number of parameter values for $(\sigma_j,\xi_j)$. Then, $\hat\psi_k=\hat\sigma_j$ and $\hat\nu_k=\hat\xi_j$ for each site $k$ in cluster $\mathcal{C}_j$. We refer to this approach as cluster-wise Monte Carlo integration (CWMC).

Return levels are an important quantity to characterize extremal behaviour. The $\tau$-year return level is the value which is exceeded on average once every $\tau$ years. For $R_{k,t}-u_k \mid (R_{k,t}>u_k)\sim \mbox{GPD}(\psi_k,\nu_k)$, the $\tau$-year return level of $R_{k,t}$ is 
\begin{equation}
u_k + \frac{\psi_k}{\nu_k}\left[\, \left( \lambda_u\,\tau\, \right)^{\nu_k} - 1\,  \right],
\label{eq:ReturnLevel}
\end{equation}
where $\lambda_u$ is the expected number of times $R_{k,t}$ exceeds $u_k$ per year. We obtain posterior median return levels similarly to the GPD parameters in \eqref{eq:MCI}. 

\section{Simulation Examples}
\label{sec:Example}

\begin{figure}
\begin{center}
\includegraphics[width=8cm, trim={5cm 5cm 5cm 5cm}]{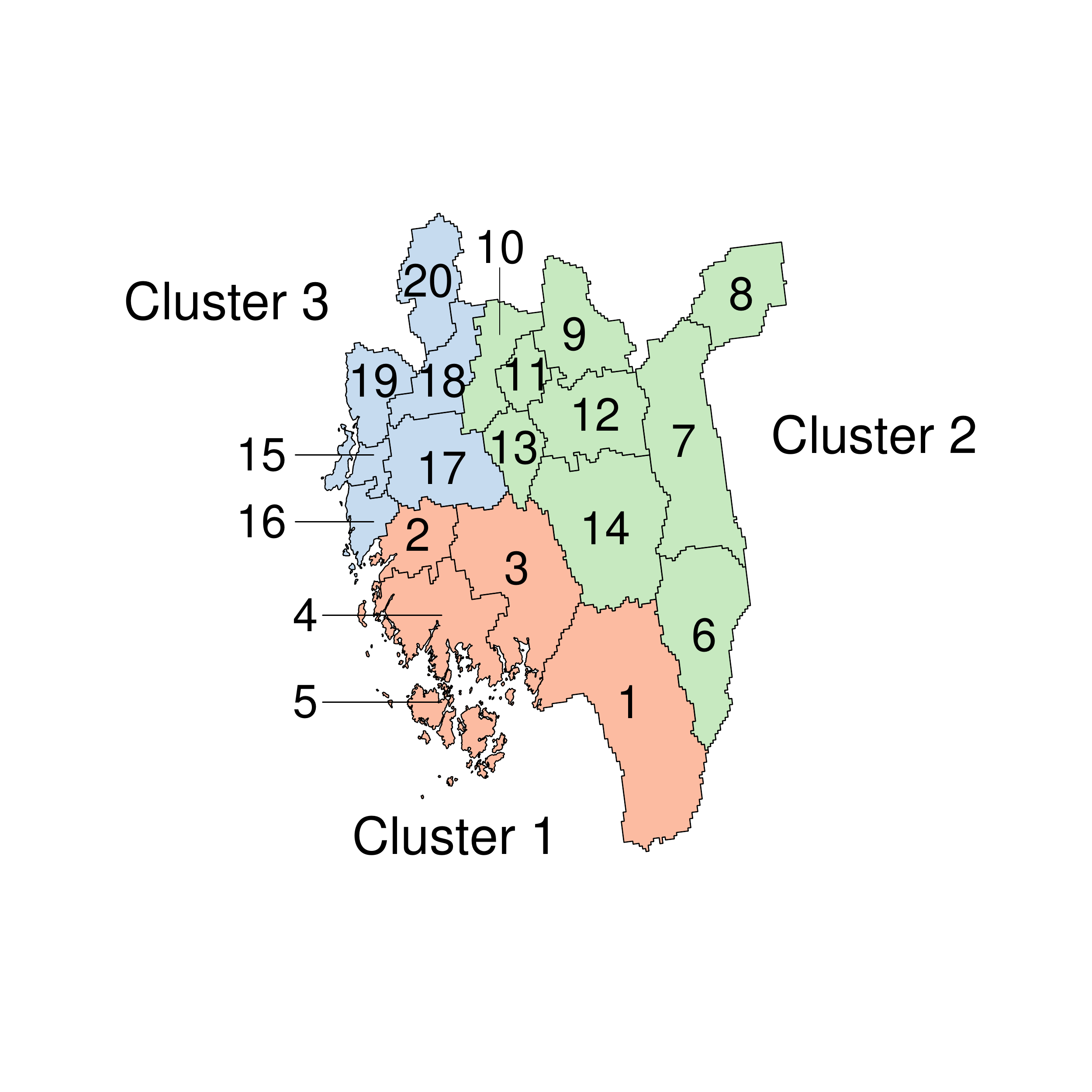}
\end{center}
\caption{Map of areal units in Section 4.}
\label{fig:MapNumericalExample}
\end{figure}

We consider the $K=20$ areal units in Figure~\ref{fig:MapNumericalExample} which correspond to a set of municipalities in south-east Norway. The distance $d_{k,k'}$~($k,k'=1,\ldots,20$) is computed using the coordinates of the municipalities' centroids and by accounting for the earth's curvature. Distances are standardized such that $0 \leq d_{k,k'} \leq 1$. The marginal distributions are set to  $R_{k,t}\sim \mbox{GPD}(\psi_k,\nu_k)$~($k=1,\ldots,20;~t=1,\ldots,100$); we set $u_k=0$ and thus obtain 100 threshold excesses per site for analysis. Data for the extremal dependence are simulated independently from the threshold excesses. We fix $Q_{k,k'}=20$~($k,k'=1,\ldots,20$) and sample $P_{k,k'}$ from the beta-binomial distribution \eqref{eq:Pkk}. Posteriors for the spatial clusters and model parameters herein are based on running the RJMCMC algorithm in Section~\ref{sec:Estimation} for $10^6$ iterations with the initial number of clusters set to $J=5$; the first $5\times10^5$ iterations are discarded as burn-in and then every 100-th sample is stored for analysis.

Our methodology aims at identifying the spatial cluster structure and estimating the marginal distributions. The simulation study illustrates the performance of our approach with respect to these key aspects. Studies~1 and 2 consider the case of all sites having the same marginal distribution and common pairwise extremal dependence structure, i.e.\ $J=1$. The studies differ in that the data are independent (strongly correlated) in Study~1 (Study~2). By investigating the case $J=1$, we check whether our approach tends to the parsimonious estimate $J=1$ or the extreme case $J=20$, i.e., each site forms its own cluster. Study~3 then considers the case of $J=3$ clusters with the cluster formulation shown in Figure~\ref{fig:MapNumericalExample}, with spatially independent and dependent data in each cluster.

\vspace{0.3cm}
\noindent {\bf Study 1:} The marginal GPD parameters are $\psi_k=2$ and $\nu_k=0.1$~($k=1,\ldots,20$), and the spatial dependence parameters are $\gamma_1=2$ and $\beta=10$; we do not specify $\gamma_0$ since $J=1$. The posterior has $J=1$ for 98\% of the sample and the point estimate for the cluster structure allocates all sites to the same cluster. The 90\% credible intervals for the site-wise GPD parameters $\psi_k$ and $\nu_k$ are $(1.9, 2.2)$ and $(0.06, 0.13)$ respectively for all sites. To illustrate the benefits of pooling information spatially, we estimate the GPD parameters for the municipality with index $k=4$ (Fredrikstad) only using the simulated threshold excesses for this site via a random walk Metropolis algorithm; this gives 90\% credible intervals of $(1.8, 2.8)$ and $(-0.06, 0.24)$ for $\psi_4$ and $\nu_4$ respectively. Consequently, our approach correctly identifies the underlying spatial cluster structure and improves estimates for the marginal distributions in this case.

\vspace{0.3cm}
\noindent {\bf Study 2:} For the data as generated in Study~1, we induce strong spatial correlation by matching ranks across sites, i.e., the $m$-th highest observation ($m=1,\ldots,100$) occurs simultaneously at all sites. The posterior probability of $J=1$ is 0.92 and we again recover the underlying cluster structure. The credible intervals for $\psi_k$ and $\nu_k$~($k=1,\ldots,20$) are $(1.5, 2.4)$ and $(-0.05, 0.28)$, and thus wider than in Study~1, as expected given there is less information in the pooled data than in Study~1 due to the strong correlation. Furthermore, the width of the credible intervals is very similar to the ones when estimating $\psi_4$ and $\nu_4$ solely based on the observations for site $k=4$ in Study 1; so we obtain almost no additional information due to the strong spatial dependence, apart from the centre of the credible interval being closer to the true value. The latter feature arises as our method pools information across the cluster while only data from site 4 is used to estimate $(\psi_4,\nu_4)$.

\begin{figure}
\begin{center}
\includegraphics[width=7.25cm]{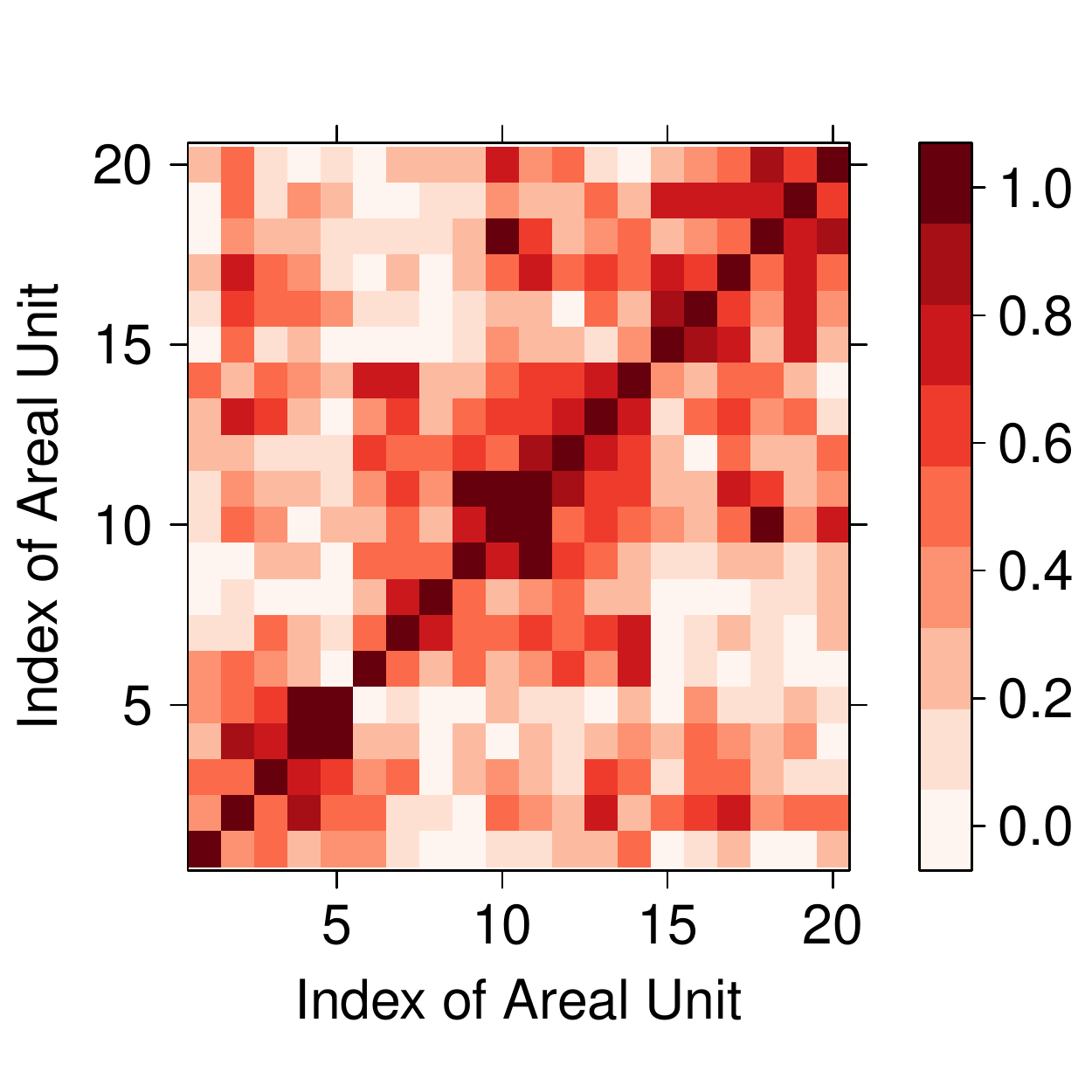}\hspace{1cm}
\includegraphics[width=7.25cm]{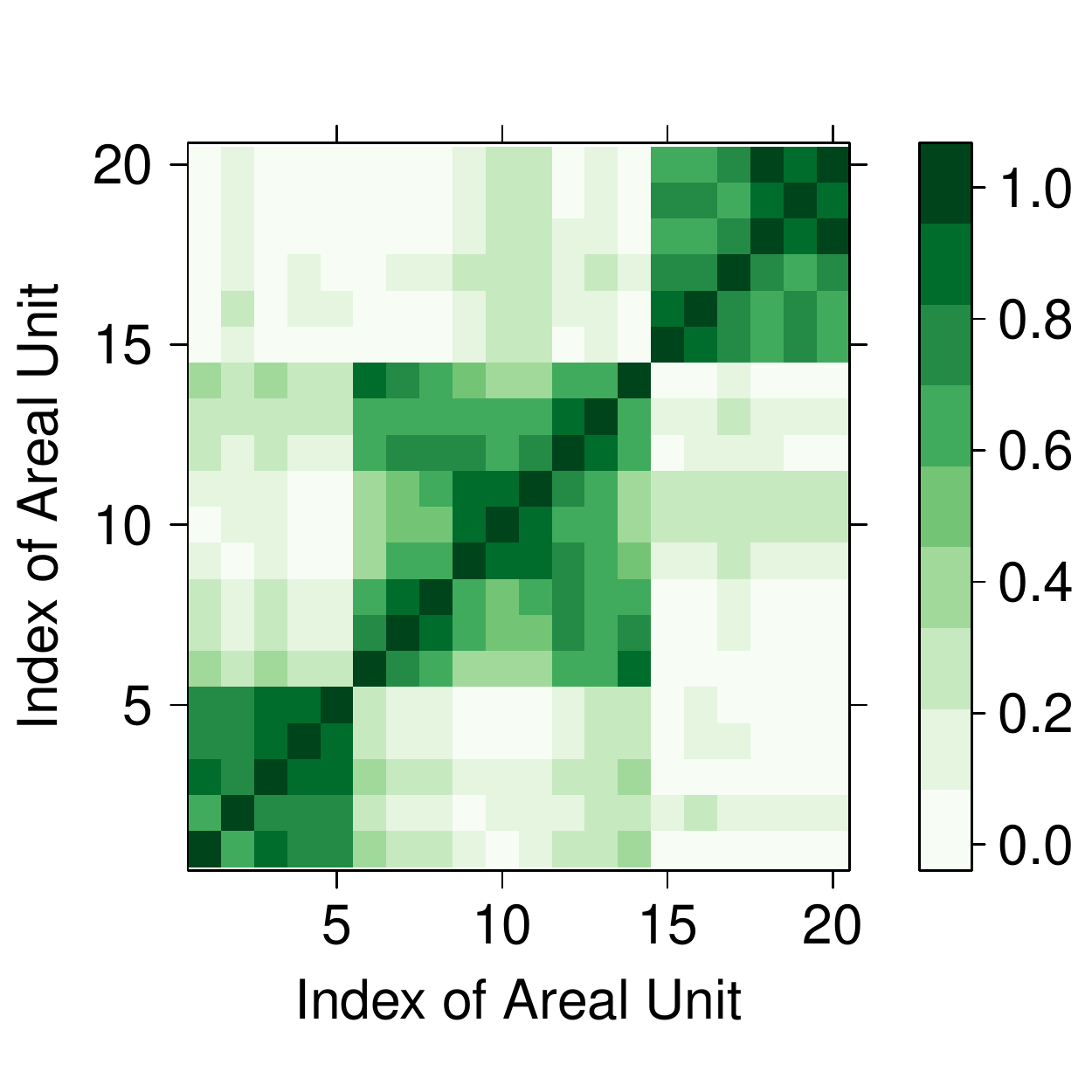}
\end{center}
\caption{Empirical estimate for the coefficients of asymptotic dependence $\chi_{k,k'}$ (left) and the posterior probability $S_{k,k'}$ of the areal units $(k,k')$ being in the same cluster (right).}
\label{fig:Sim3}
\end{figure}

\vspace{0.3cm}
\noindent {\bf Study 3:} We simulate threshold excesses independently across sites with the cluster-specific GPD parameters $\bm\sigma = (2.0,2.3,2.6)$ and $\bm\xi = (0.05,0.10,0.15)$. The dependence parameters are $\gamma_0=3$, $\gamma_j=2~(j=1,2,3)$ and $\beta=10$; Figure~\ref{fig:Sim3} left panel shows the empirical estimates $\hat\chi_{k,k'}=P_{k,k'}/Q_{k,k'}$~($k,k'=1,\ldots,20$) for $\chi_{k,k'}$. The matrix $\mathbf{S}$, given by expression~\eqref{eq:SimilarityMatrix}, shown in Figure~\ref{fig:Sim3} right panel indicates that $J=3$ clusters are present; $\mathbf{S}$ has the form of a block diagonal matrix with each block only containing entries of high probability. One may conclude that, for instance, the sites 1-5 form a cluster; this would be correct given Figure~\ref{fig:MapNumericalExample}. Indeed, the point estimate for the spatial clusters is identical to the true cluster formulation in Figure~\ref{fig:MapNumericalExample}. The 90\% central credible interval for the number $J$ of clusters is (2,7), and the true marginal GPD parameters lie within the estimated 90\% central credible interval for all sites. We then generate dependent data using a Gaussian copula and the GPD marginals above. Again, the spatial cluster structure is correctly identified and the spatial dependence leads to a higher uncertainty on the site-wise GPD parameters (results not shown). 

\section{Data analysis}
\label{sec:Application}

We now apply our methodology to analyze the data described in Section~\ref{sec:Data}. After an appropriate burn-in period, we perform $1.5\times10^7$ iterations of the RJMCMC sampling scheme in Section~\ref{sec:Estimation}, with every 1000-th sample being stored. Initially, 10\% of the sites are cluster centres. The acceptance probabilities for birth (and death) were 0.02 and 0.05 for the data in Sections \ref{sec:ResultsNorway} and \ref{sec:ResultsUK} respectively. Our \texttt{C++} implementation took about 40 minutes per $10^6$ iterations on a standard laptop for the larger data set in Section~\ref{sec:ResultsNorway}. 
 
\subsection{Daily precipitation in South Norway}
\label{sec:ResultsNorway}

Since an extreme event may be split across two days, we consider the aggregated amount of precipitation over a 48-hour period. Further, we decluster by using only the highest observation per week because consecutive observations are strongly correlated; this gives about 500 observations per municipality. We then select a threshold $u_k$~($k=1,\ldots,343$) for which $R_{k,t}-u_k\mid R_{k,t}>u_k$ approximately follows a GPD. Analysis of the threshold stability plots suggests to set $u_k$ to the empirical 92.5\% quantile; this gives about 40 peaks-over threshold per municipality. The empirical measures $P_{k,k'}$ and $Q_{k,k'}$~($k,k'=1,\ldots,343$) are derived based on the dependence threshold $\tilde{u}=0.95$. 

\begin{figure}
    \centering
    \includegraphics[width=0.48\textwidth]{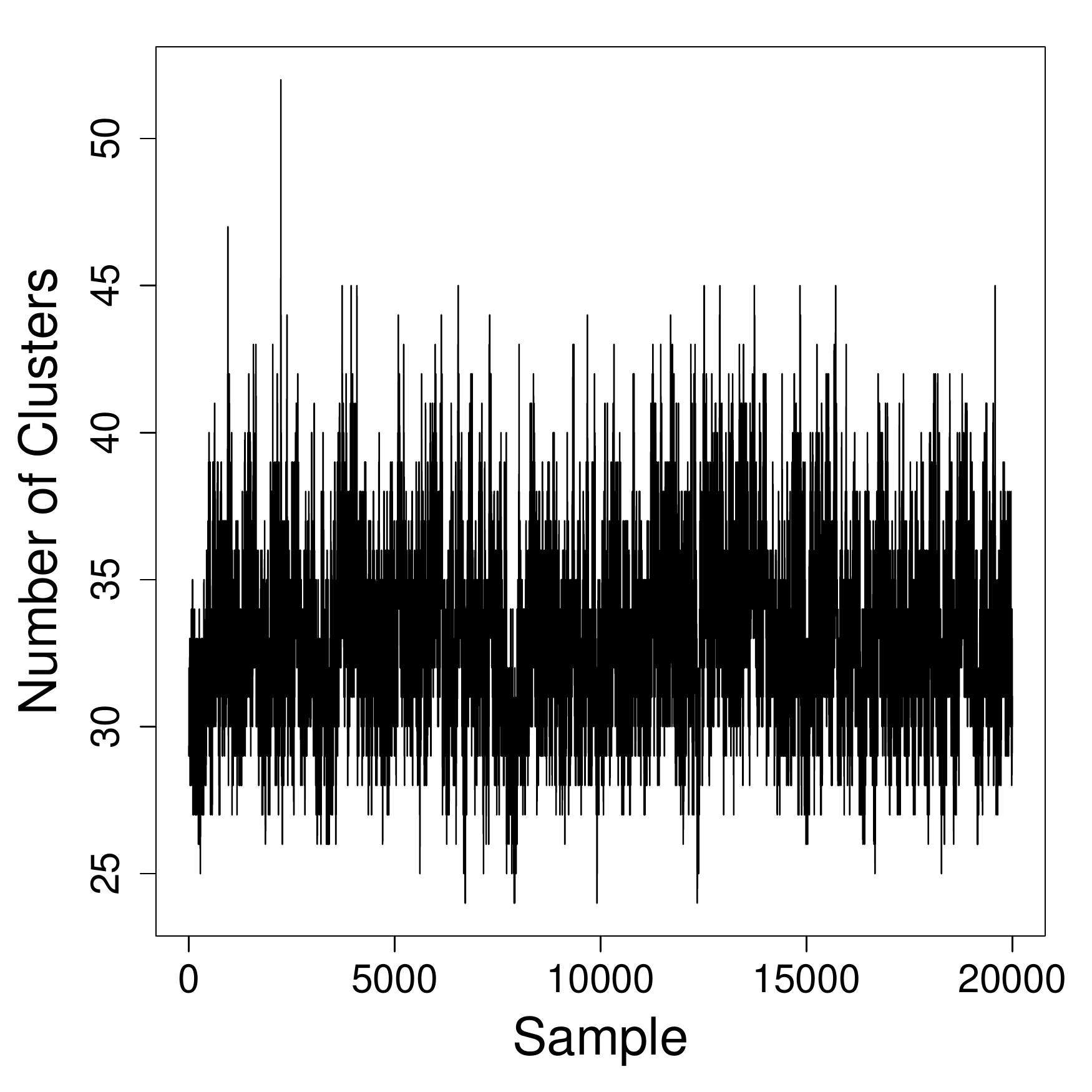}
    \hspace{0.3cm}
    \includegraphics[width=0.48\textwidth]{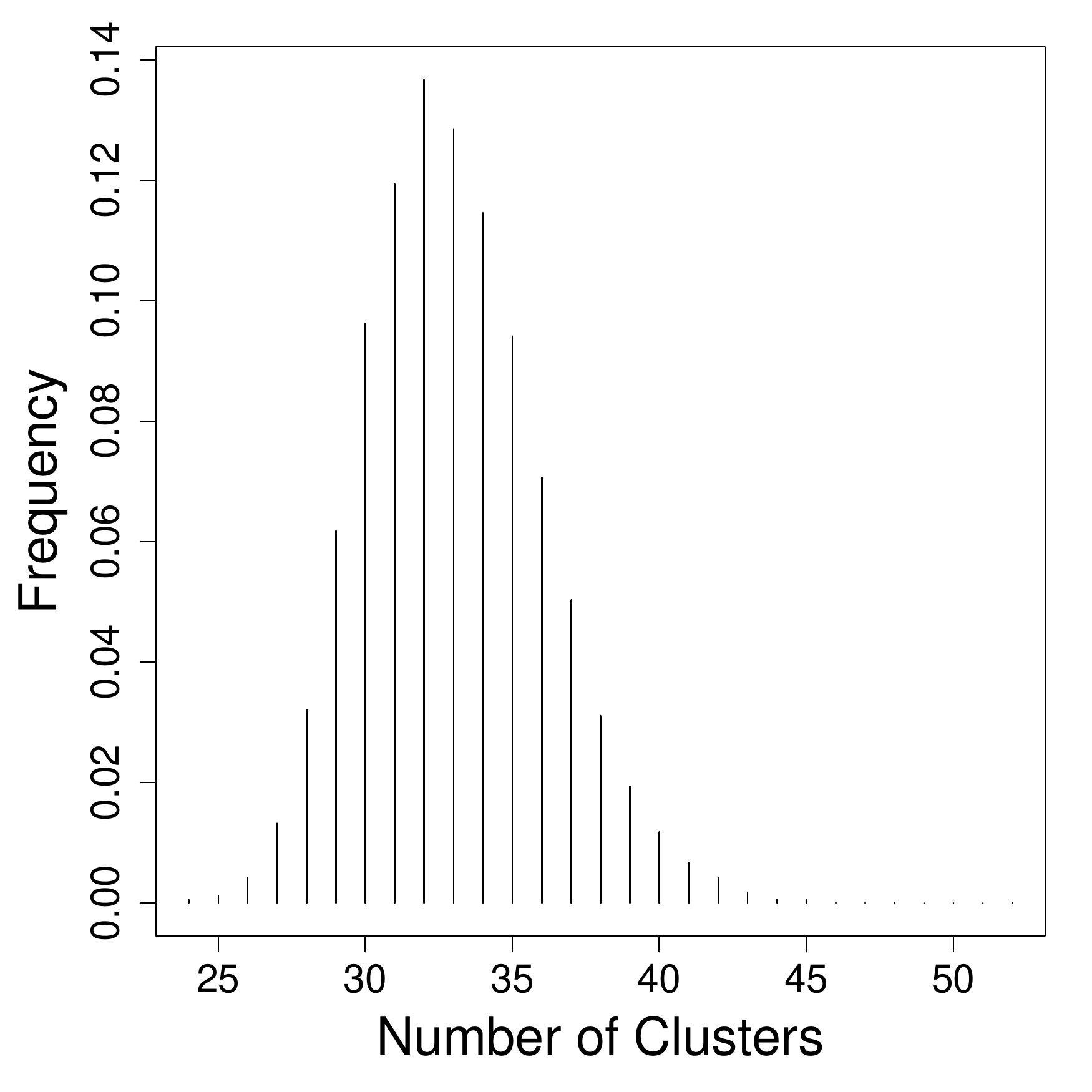}
    \caption{Trace plot (left) and posterior mass function (right) of the number $J$ of clusters for the precipitation data in Section~\ref{sec:ResultsNorway}.}
    \label{fig:NumberClustersNorway}
\end{figure}

\begin{figure}
    \centering
    \includegraphics[width=0.48\textwidth]{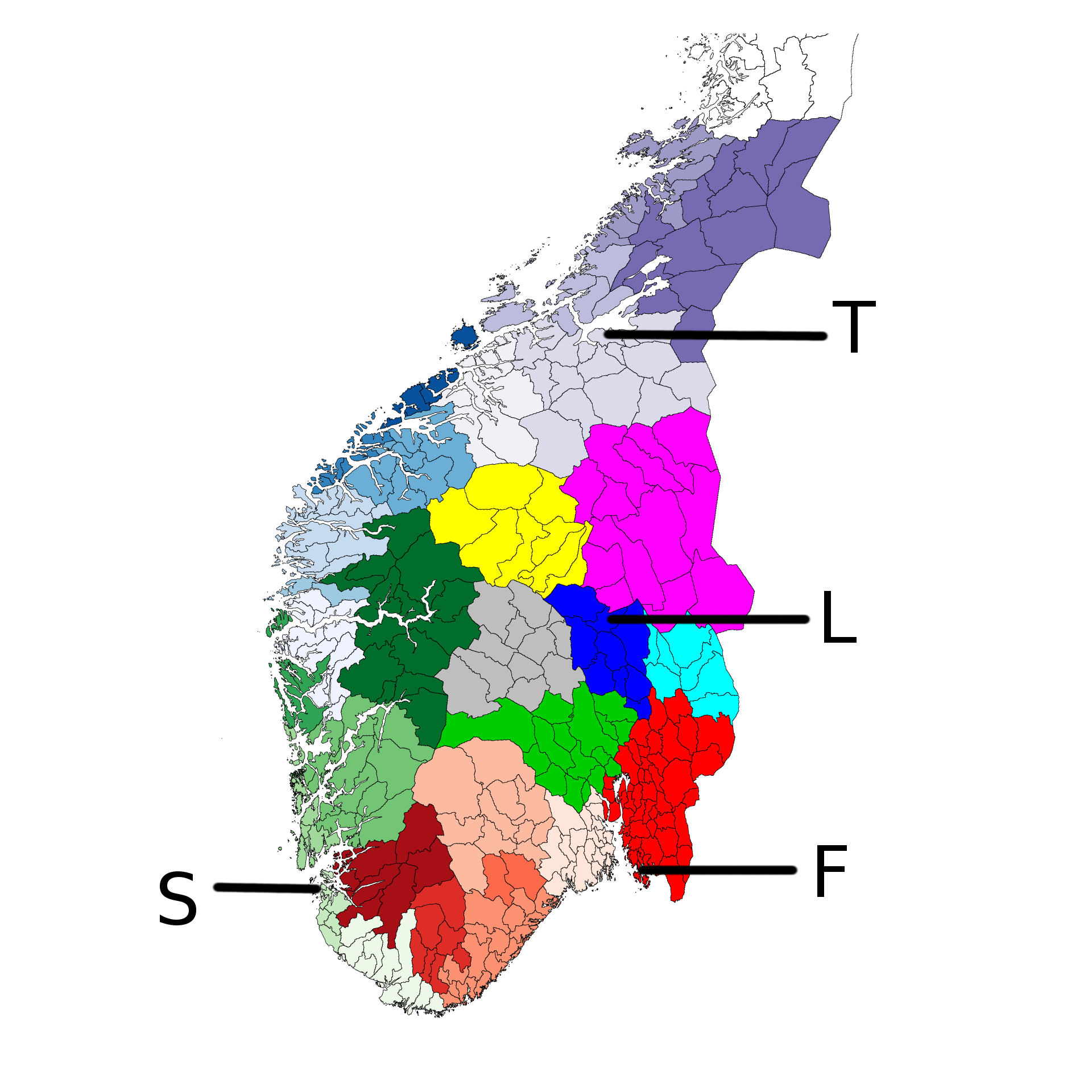}
    \hspace{0.3cm}
    \includegraphics[width=0.48\textwidth]{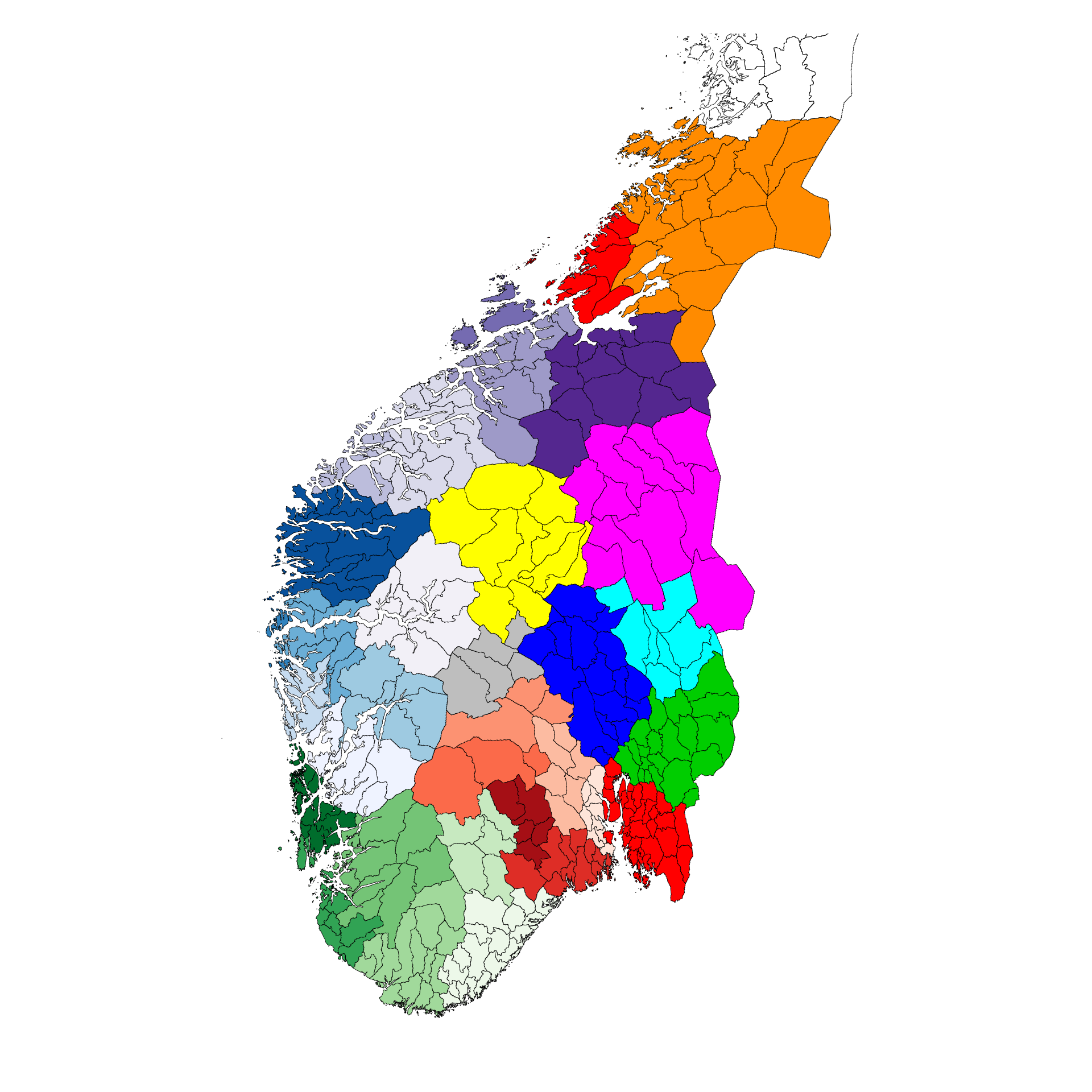}
    \caption{Point estimate of the spatial cluster structure for the Norwegian precipitation data in Section~\ref{sec:ResultsNorway}: when using peaks-over threshold and extremal dependence (left) and when only using peaks-over threshold (right). The municipalities Fredrikstad (F), Lillehammer (L), Stavanger (S) and Trondheim (T) are highlighted in the left panel.}
    \label{fig:ClustersNorway}
\end{figure}

Figure~\ref{fig:NumberClustersNorway} left panel shows appropriate convergence and mixing of the sampled Markov chain for the number of clusters. The posterior mean number of clusters is $J=33$ and the posterior mass function of $J$ is shown in Figure~\ref{fig:NumberClustersNorway} right panel; the central 80\% credible interval is $(29,37)$. The point-estimate for the spatial cluster structure comprises 30 clusters which are illustrated in Figure~\ref{fig:ClustersNorway} left panel. Most clusters contain between 7 and 16 municipalities; two clusters located in western and southern Norway contain only two municipalities while the cluster in the south-east includes more than 40, mostly small, areal units. The derived spatial clusters agree with known climatology. Clusters along the west coast regularly observe very high amounts of precipitation which are often related to the Gulf Stream. We also find that the drier municipalities in central Norway are grouped together. 

To investigate the effect of exploiting information on both marginal tail behaviour and extremal dependence, we consider a second spatial clustering algorithm with likelihood function $L_{\mathrm{M}}(\bm\theta_{\mathrm{M}}^{(J)}\mid\mathcal{D},\mathbf{Z})$, i.e., clusters are solely based on threshold exceedances while spatial dependence is ignored, except via the adjustment \eqref{eq:LikeAdj}. The sampled number of clusters is higher than for our approach; the posterior mean for $J$ is 39 with 80\% credible interval (33,45). Figure~\ref{fig:ClustersNorway} right panel shows the point-estimate for the spatial clusters; seven clusters contain five or less municipalities while three clusters include more than 20 municipalities. The highest observations per municipality for the most northern cluster in Figure~\ref{fig:ClustersNorway} ranges from 57.3mm through to 174.2mm in comparison to (94.5mm, 174.2mm) for our approach. As such, our proposal to incorporate both marginal distributions and extremal dependence seems to produce more realistic clusters than the considered alternative. 

\begin{table}
    \centering
    \begin{tabular}{r c c c c c}
    \vspace{-0.3cm}~\\\hline    \vspace{-0.3cm}~\\

    $\tau$  &  Method & Fredrikstad  & Lillehammer   & Stavanger      & Trondheim    \vspace{0.3cm}\\
    \hline \vspace{-0.3cm}~\\

    25      &  (i)    & 72~(60, 99)  & 74~(64, 105)  & 96 ~(81, 133)  & 94~(78, 135) \\
            &  (ii)   & 72~(65, 82)  & 82~(68,  99)  & 103~(87, 127)  & 90~(79, 106) \\
            &  (iii)  & 69~(63, 75)  & 81~(68,  97)  & 99 (83, 130)   & 91~(80, 108) \vspace{0.2cm}\\
        
    100     &  (i)    & 87~(68, 143) & 81~(66, 130)  & 125~(90, 191)  & 112~(87, 188) \\
            &  (ii)   & 82~(73, 101) & 98~(78, 126)  & 131~(101, 172) & 112~(94, 143) \\
            &  (iii)  & 80~(71, 86)  & 95~(78, 120)  & 123~(95, 191)  & 114~(95, 149) \vspace{0.3cm}\\
    \hline
    \end{tabular}
       \caption{Posterior median (central 90\% credible interval) of the $\tau$-year return level in mm for four municipalities with $\tau=(25,100)$. The considered methods are (i) Individual estimation for each municipality (ii) SWMC and (iii) CWMC.}
    \label{tab:ReturnLevelsNorway}
\end{table}

We conclude the analysis by estimating return levels for the four municipalities highlighted in Figure~\ref{fig:ClustersNorway}: Fredrikstad (F), Lillehammer (L), Stavanger (S) and Trondheim (T). The selected municipalities differ in both climate and clustering; Fredikstad and Trondheim lie in large clusters while Stavanger and Lillehammer are members of relatively small clusters. Since $u_k$ is the 92.5\% quantile, the average number of exceedances per year in \eqref{eq:ReturnLevel} is $\lambda_u = 52(1-0.925) = 3.9$. We consider three approaches to estimate return levels. For the first estimate, we obtain return level estimates individually using only the observed peaks-over threshold of the municipality, i.e., estimates are based on the 40 observed threshold excesses for the municipality. Our other estimates are based on SWMC and CWMC as described in Section~\ref{sec:Pointestimate}. 

Table~\ref{tab:ReturnLevelsNorway} shows that SWMC and CWMC provide shorter credible intervals for the considered municipalities than method (i); the central 90\% credible interval for both method (ii) and (iii) always lies within the credible interval of method (i). The credible intervals obtained by SWMC and CWMC are very similar for Lillehammer and Trondheim; this indicates that our point estimate for the cluster structure groups municipalities which have indeed very similar marginal distributions. We further see that CWMC produces shorter credible intervals than SWMC for Fredrikstad; this arises as the SWMC approach accounts for cluster uncertainty. In the CWMC method, the cluster containing Fredrikstad contains many municipalities and spatial variation in the GPD parameters is oversmoothed due to the cluster structure being fixed; this can be seen in Table~\ref{tab:ReturnLevelsNorway} since the credible interval for method (iii) does not contain the posterior median derived by method (i). The SWMC approach, on the other hand, performs better since the municipalities within this cluster are regularly split across smaller clusters; this allows for a good exploration of the parameter space. Finally, the CWMC credible interval is similar to method (i) for Stavanger while SWMC gives a narrower credible interval. The cluster containing Stavanger is relatively small in size and an extreme event usually affects most municipalities within the cluster. Due to this strong spatial correlation, we obtain little additional information by pooling information across the fixed cluster. Conversely, our SWMC approach efficiently pools information from a larger set of municipalities because the cluster structure is allowed to change. 

\subsection{Daily river flow in the UK}
\label{sec:ResultsUK}

The data described in Section~\ref{sec:DataRF} exhibit a strong seasonal pattern for all $K=45$ gauges. We consider separately the maximum weekly river flow for November-March and May-September for which in each case the assumption of stationarity seems reasonable. The observations in each season are then standardized site-wise to mean 0 and variance 1; while this affects the scale parameter $\psi_k$ of the GPD in \eqref{eq:introGPD}, it is well known that this leaves the shape parameter $\nu_k$ unchanged.

A common threshold across all gauges, $u_1=u_2\cdots=u_K$, is selected individually for the two seasons, giving between 25 and 40 peaks-over threshold per season for most sites. The threshold $\tilde{u}$ in \eqref{eq:ChiEmpirical} is set to $\tilde{u} = 0.96$ for November-March and $\tilde{u}=0.95$ for May-September. This gives $Q_{k,k'}\approx 35$ (25) for most pairs of sites $(k,k')$~($k,k'=1,\ldots,45$) over the summer (winter) seasons. Here, the distance $d_{k,k'}$ between sites $(k,k')$ is set to their hydrological distance (Section 2) which we scale such that $0 \leq d_{k,k'}\leq 1$. 

\begin{figure}
    \centering
    \includegraphics[width=0.45\textwidth]{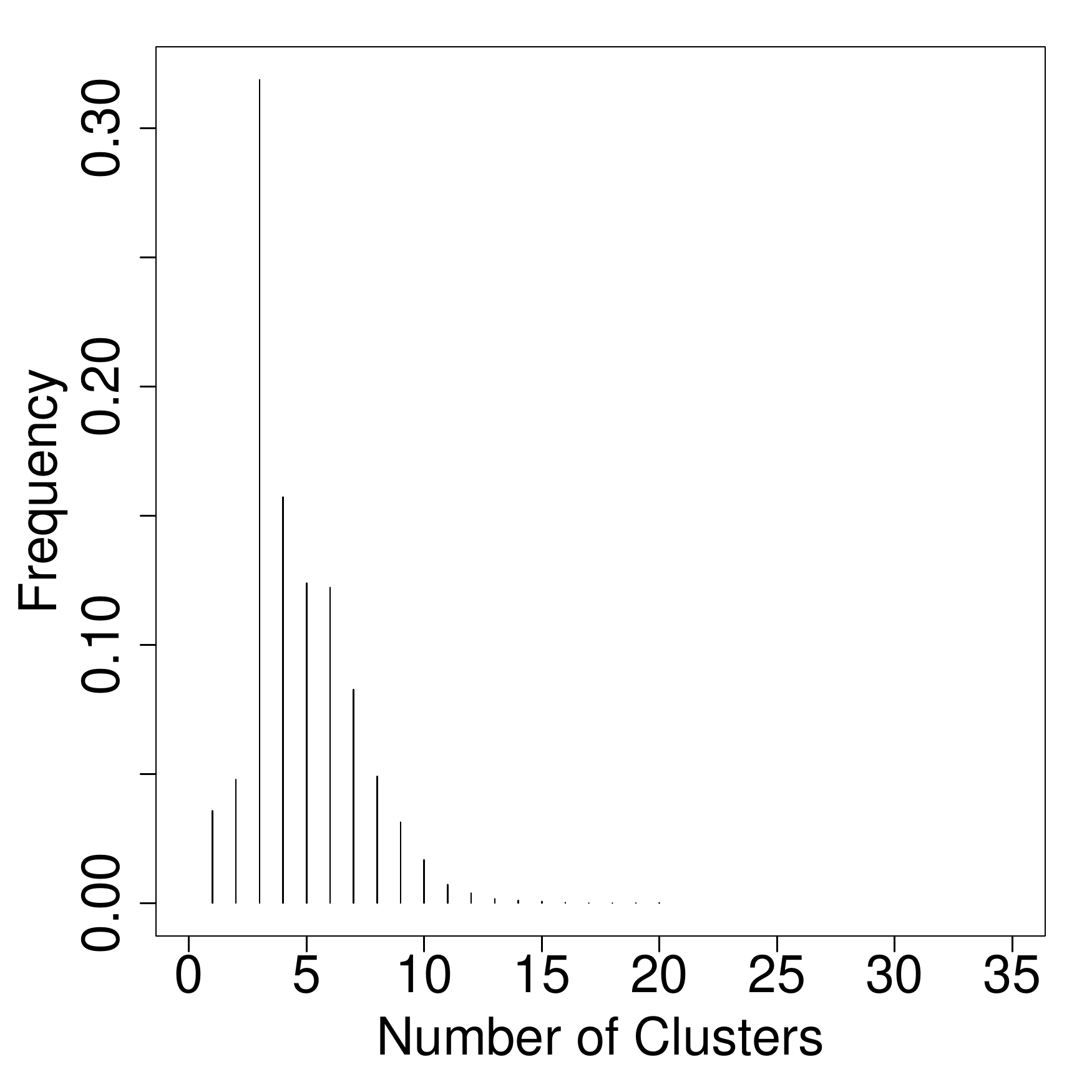}
    \hspace{1cm}
    \includegraphics[width=0.45\textwidth]{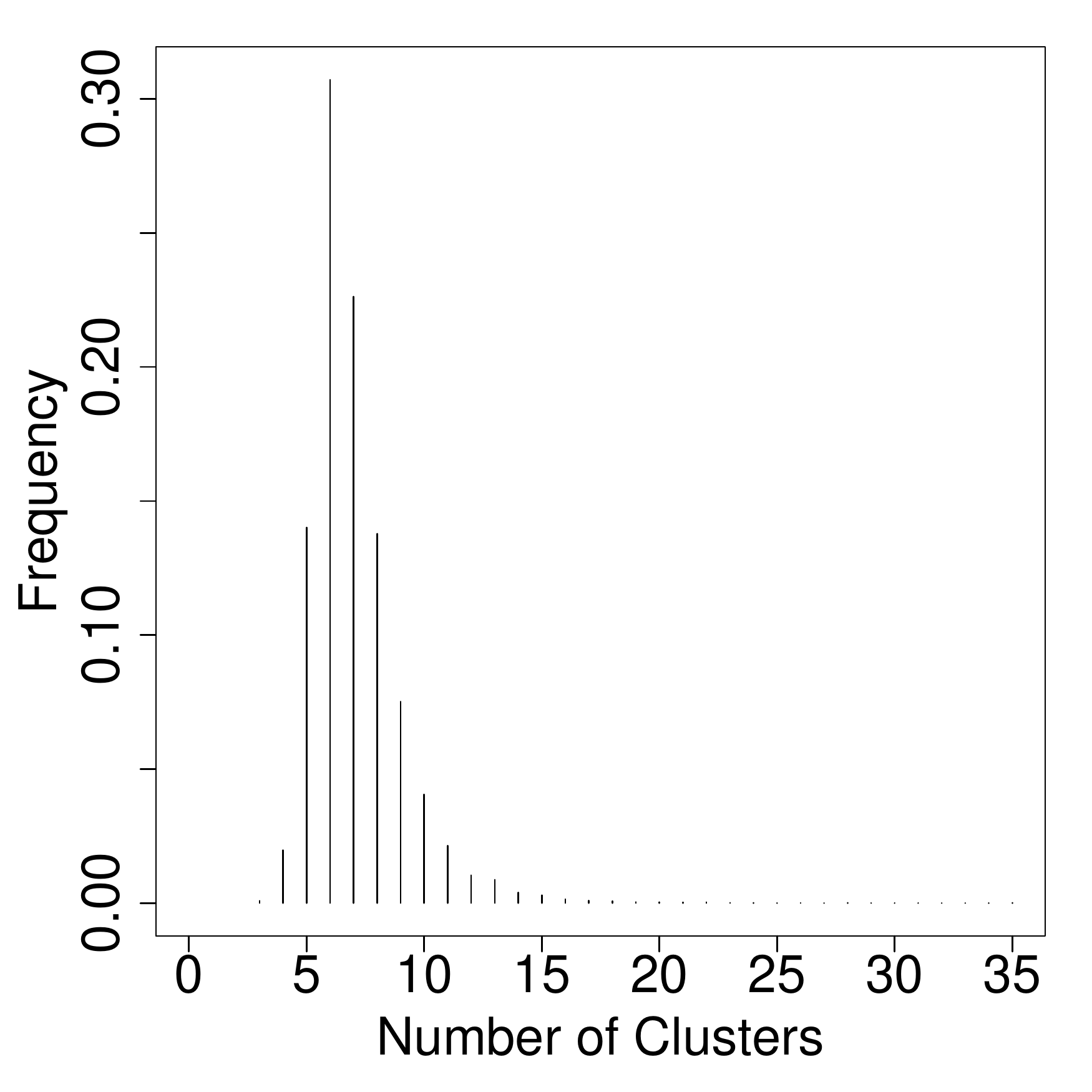}
    \caption{Posterior mass functions of the number $J$ of clusters for November-March (left) and May-September (right).}
    \label{fig:NumberClusters}
\end{figure}

The posterior mass functions for the number $J$ of clusters for the two seasons in Figure~\ref{fig:NumberClusters} indicate that the number of clusters is higher for May-September than for November-March; the 80\% credible intervals are (5,9) and (3,8). This result agrees with known climatology. Extreme river flows in winter in the UK are often caused by extratropical cyclones which affect larger areas; we thus expect larger clusters for November-March than for May-September. 

\begin{figure}[t!]
    \centering
    \includegraphics[width=0.45\textwidth]{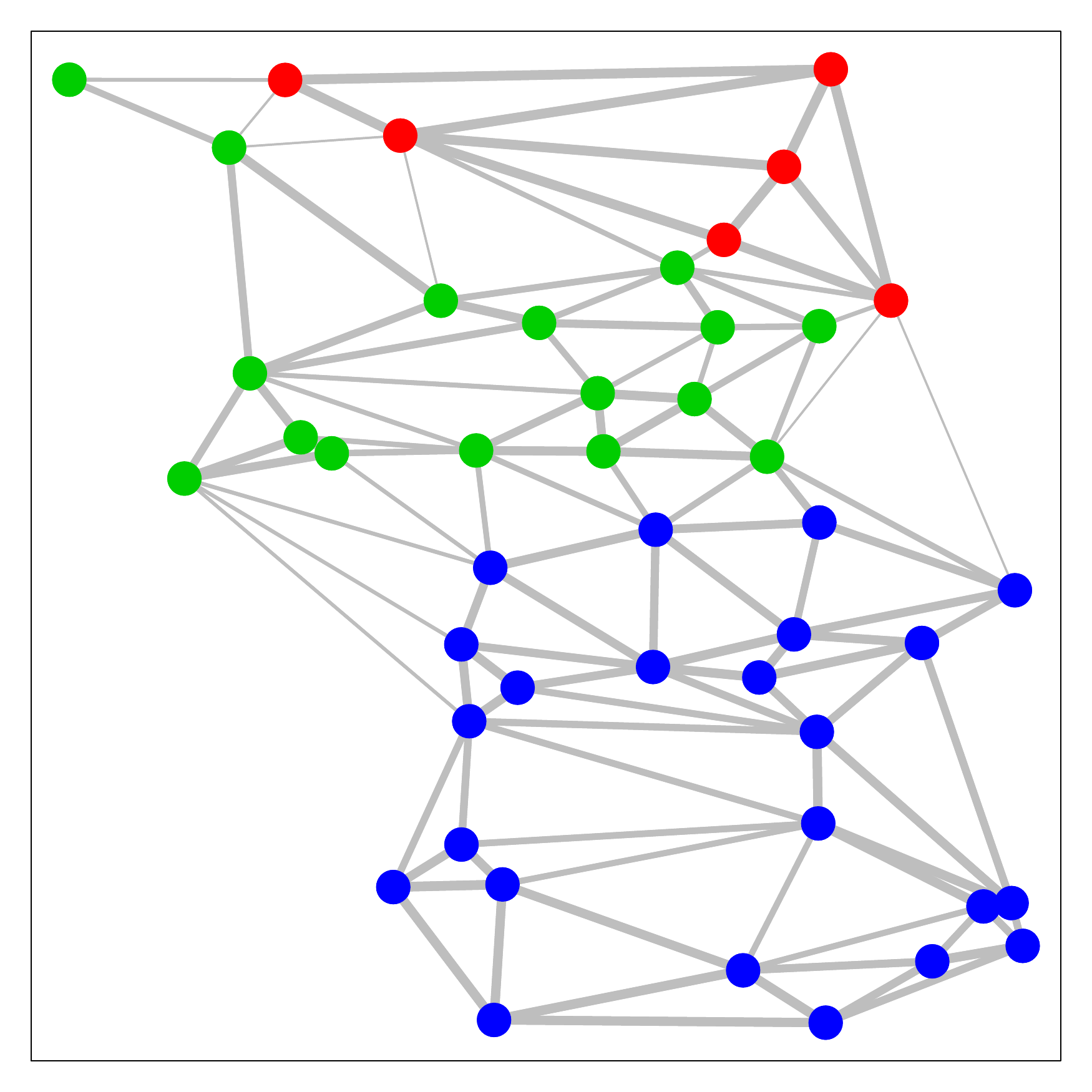}
    \hspace{1cm}
    \includegraphics[width=0.45\textwidth]{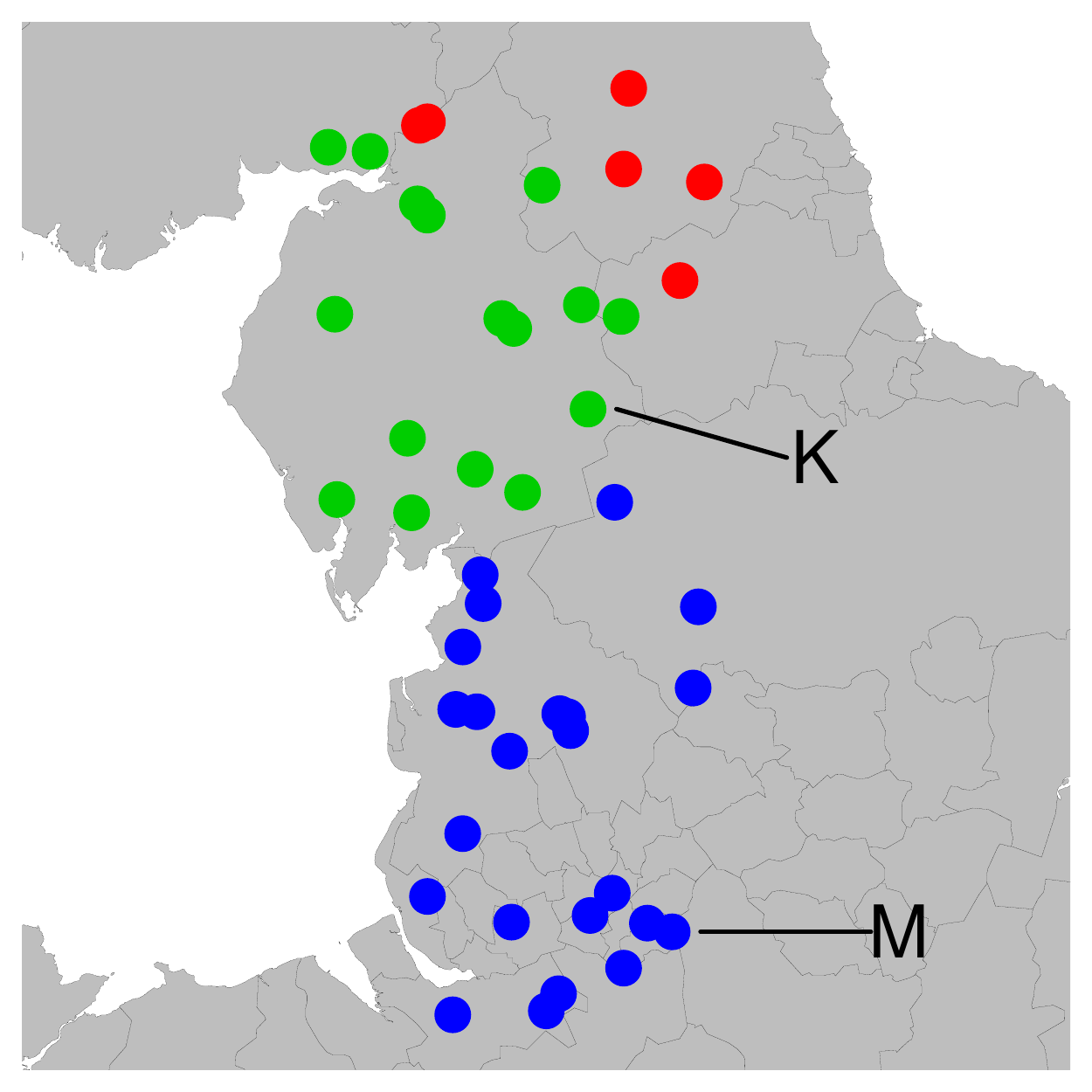}\\~\\
    \includegraphics[width=0.45\textwidth]{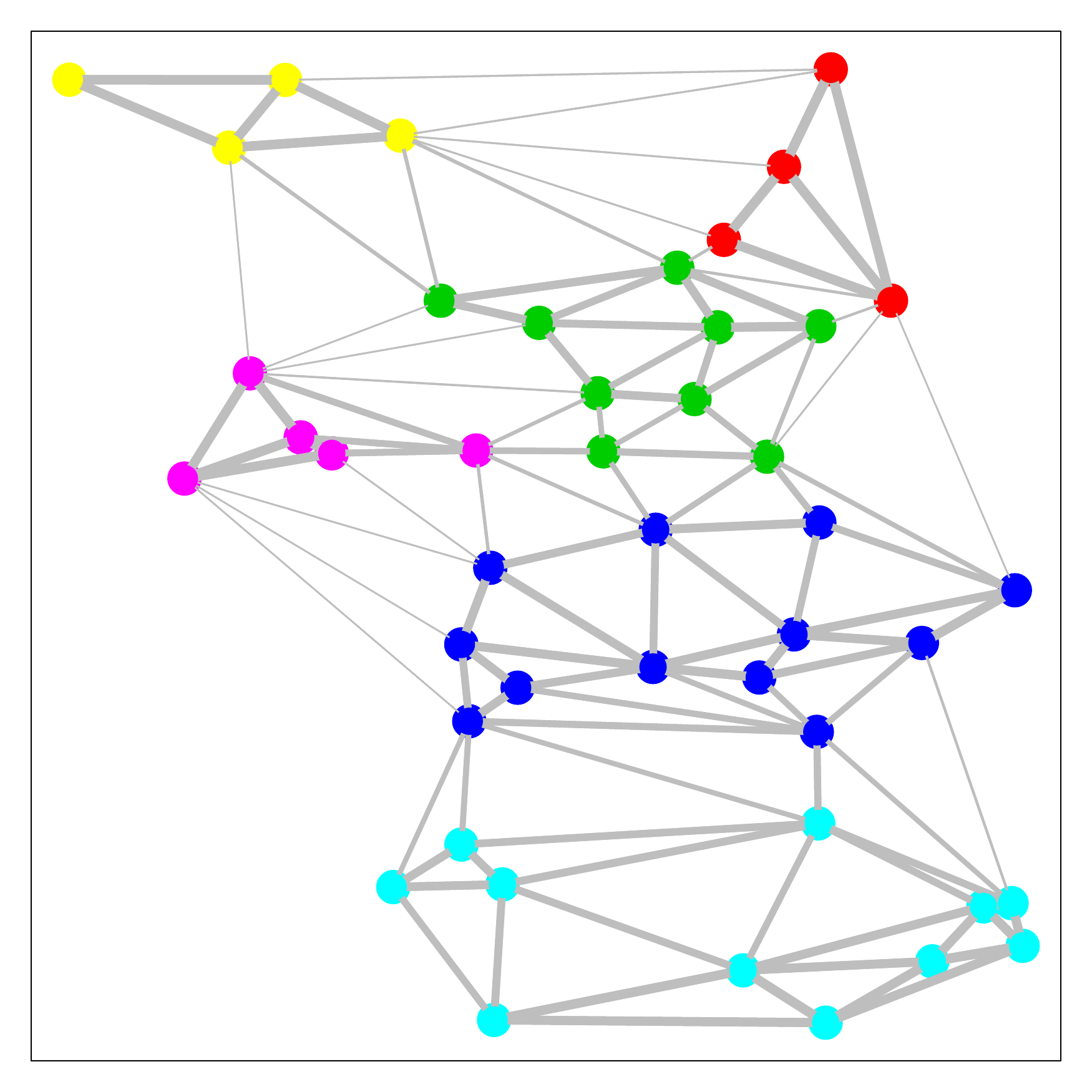}
    \hspace{1cm}
    \includegraphics[width=0.45\textwidth]{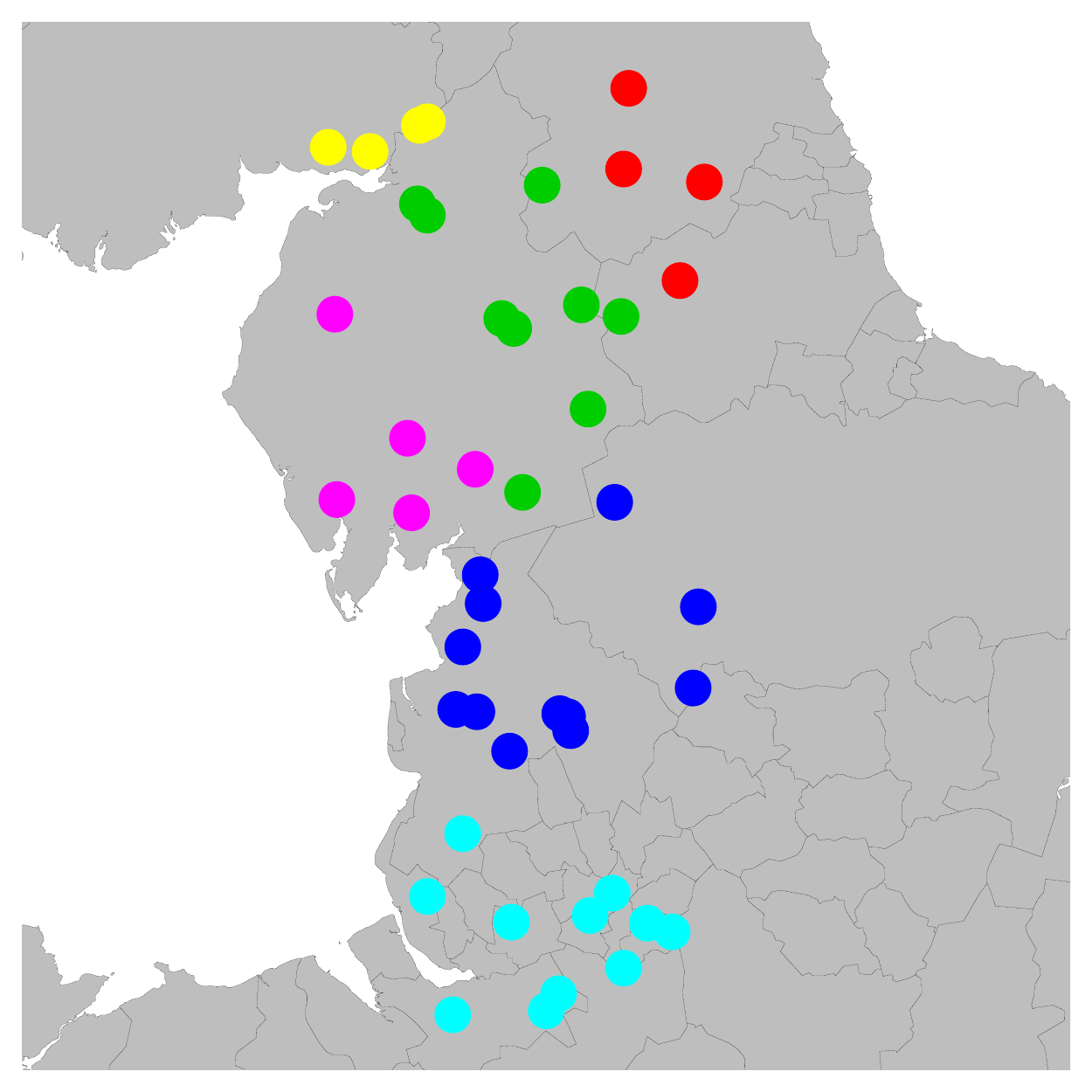}
    \caption{Point estimates for the underlying spatial cluster structure for November-March (top) and May-September (bottom). The left plots show the cluster allocation on the space of hydrological coordinates. A line between two sites indicates that they are considered adjacent and the line width corresponds to the posterior probability of them being in the same cluster. The right plots illustrate the derived clusters with respect to their latitude and longitude coordinates. The solid black lines are the boundaries of the metropolitan and non-metropolitan counties, and the highlighted gauges are Kirkby Stephen (K) and Marple Bridge (M).}
    \label{fig:RiverFlowClusters}
\end{figure}

Figure~\ref{fig:RiverFlowClusters} left panels show the imposed adjacency structure and the derived seasonal clusters with respect to the hydrological coordinates of the gauges. The width of the lines in the plots further illustrates that some gauges, which are allocated to different clusters in the point estimate, are also often grouped in the same cluster; see, for instance, the central and southern cluster for November-March. We see an interesting match between the point estimates of the underlying cluster structure for the two seasons. The southern cluster for November-March splits into two equally-sized clusters in May-September. Similarly, the other large cluster (green) for the winter season splits into two clusters in summer. 

We then consider the geographical locations in Figure~\ref{fig:RiverFlowClusters} right panels for further analysis. Beyond the seasonal pattern of the clusters, it is also interesting to investigate how these compare to the underlying river networks; the estimated clusters are in fact not identical to the river networks. For instance, we have five gauges for the River Tyne in North-East England but only four of these are allocated to the same cluster (the red cluster for the summer season). Finally, we note that some clusters are split along administrative boundaries, in particular, in winter.

\begin{table}
    \centering
    \begin{tabular}{rccccc}
    \hline \vspace{-0.3cm}\\
    $\tau$ & Model & \multicolumn{2}{c}{Kirkby Stephen} & \multicolumn{2}{c}{Marple Bridge} \\
           &       & Winter          & Summer           &   Winter     &  Summer     \vspace{0.3cm}\\
    \hline\vspace{-0.3cm}~\\
    
    100    & (i)   & 107 (85, 164)    & 45 (36, 70)     & 70 (55, 111) & 34 (26, 52) \\
           & (ii)  & 98  (84, 131)    & 45 (37, 57)     & 65 (55, 92)  & 36 (32, 43) \\
           & (iii) & 110 (86, 153)    & 49 (43, 60)     & 60 (55, 69)  & 36 (32, 43) \\
            ~\vspace{-0.3cm}\\
           
    500    & (i)   & 138 (99, 269)    & 56 (41, 108)    & 95 (67, 196) & 45 (31, 90) \\
           & (ii)  & 128 (103, 200)   & 57 (43, 80)     & 84 (67, 148) & 47 (39, 64) \\
           & (iii) & 157 (110, 266)   & 63 (52, 85)     & 77 (67, 94)  & 47 (37, 66) \\
            ~\vspace{-0.3cm}\\
    \hline
    \end{tabular}
    \caption{Posterior median (central 90\% credible interval) of the $\tau$-year return level in m$^3$/s for two gauges with $\tau=(100,500)$. The considered methods are (i) Individual estimation for each municipality (ii) SWMC and (iii) CWMC.}
     \label{tab:ReturnLevelsUK}
\end{table}

To conclude, we estimate return levels for $\tau=(100,500)$ for the two gauges highlighted in Figure~\ref{fig:RiverFlowClusters} top-right panel: Kirkby Stephen (K) and Marple Bridge (M). Kirkby Stephen was selected because multiple of its adjacent sites are allocated to different clusters, while Maple Bridge lies centrally in the southern cluster in winter and at the edge of the most southern cluster in the summer season. We consider the same inference approaches as in Section~\ref{sec:ResultsNorway}. Table~\ref{tab:ReturnLevelsUK} shows that SWMC again produces narrower credible intervals than model (i). Furthermore, the weaknesses of the CWMC approach described in Section~\ref{sec:ResultsNorway} affect the estimates strongly, in particular, for Marple Bridge and winter. 

\section{Discussion}
\label{sec:Discussion} 
We introduced a Bayesian clustering approach which groups geographical sites based on their marginal tail behaviour and their extremal dependence. The likelihood for the peaks-over threshold accounts for the spatial dependence usually found in hydrological applications. Our model for the extremal dependence postulates that sites within the same cluster exhibit higher pairwise extremal dependence than sites in different clusters, and that the degree of dependence decreases with an increasing distance between sites. Clusters are represented by their centre, which imposes them to be contiguous and leads to a good computational performance. Samples from the posterior distribution were obtained using a reversible jump MCMC algorithm. A point-estimate for the spatial cluster structure was derived from the pairwise posterior probabilities of sites being in the same cluster using Bayesian decision theory. 

We applied our approach to analyze precipitation levels across South Norway and the derived clusters agree with climatology; these clusters will be used to analyze the association between weather events and property insurance claims. The cluster approach was further applied to river flow data in the UK. We found that clusters are not identical to the river networks and that the spatial extent of extreme river flow levels is larger over winter than summer. The results also showed that our approach efficiently pools spatial information to impoove return level estimates.       

There are various ways to extend the model presented in this paper. Firstly, we model extremal dependence of a pair of adjacent sites $(k,k')$ in different clusters via the single parameter $\gamma_0$. Instead of $\gamma_0$ being constant in \eqref{eq:ChiCluster}, $\gamma_0$ may be defined as a function of the cluster-specific parameters. Consider a pair of adjacent sites $(k,k')$ with $Z_k=j$ and $Z_{k'}=j'$, $j\neq j'$. High values of $\gamma_j$ and $\gamma_{j'}$ may then imply a high value for $\gamma_0$. Another possible extension is the consideration of temporal variations in the distribution of the peaks-over threshold and/or the extremal dependence. Such an extension should then also allow for a potential change of the spatial cluster structure across seasons; the results for the UK river flow data indicate the presence of such a temporal variation in the number of clusters. 

\section*{Acknowledgement}
Christian Rohrbeck is beneficiary of an AXA Research Fund postdoctoral grant. We gratefully acknowledge funding from the King Abdullah University of Science and Technology (KAUST) Office of Sponsored Research (OSR) project "Statistical Estimation and Detection of Extreme Hot Spots with Environmental and Ecological Applications" (Award No.\ OSR-2017-CRG6-3434.02). We thank Rob Lamb, Raph{\"e}l Huser and Daniel Cooley for helpful comments and suggestions, and Ida Scheel and Ross Towe for providing access to the Norwegian rainfall data and UK river flow data respectively.

\appendix

\section{Details of the reversible jump MCMC algorithm}
\label{sec:RJMCMC}
\subsection*{Birth and death}
Suppose we currently have $J$ clusters with centres $\mathbf{C}^{(J)}\subseteq\left\{1,\ldots,K\right\}$ and parameters $\left(\bm\theta_{\mathrm{M}}^{(J)}, \bm\theta_{\mathrm{D}}^{(J)}\right)$. In a birth, we first uniformly sample a new cluster centre $C^*$ together with the index $j^*$ at which to insert it in the set $\mathbf{C}^{(J)}$; the probabilities are $\mathbb{P}\left(C^*=k \mid \mathbf{C}^{(J)}\right) = (K-J)^{-1}~(k\in\left\{1,\ldots,K\right\}\setminus\mathbf{C}^{(J)})$ and $\mathbb{P}(j^*\mid \mathbf{C}^{(J)}) =(J+1)^{-1}$~($j^*=1,\ldots,J+1$). For the new set of cluster centres, $\mathbf{C}^* = (C_1,\ldots,C^*_{j^*},\ldots,C_{J+1})$, we then derive the cluster labels $\mathbf{Z}^*$ according to \eqref{eq:Clustering}. 

To complete the proposal, we sample the parameters $(\epsilon^*,\sigma^*,\xi^*)$ of the new cluster. The mean of each proposal distribution is set to the current average value of that parameter across the sites allocated to the new cluster, while the variance of the proposal is set to the one of the corresponding prior. Let $\mathcal{C}^* = \left\{k:Z_k^*=j^*\right\}$ denote the sites allocated to the new cluster. The proposal $\xi^*$ is sampled from a normal distribution with 
\[
\xi^*\mid \left(\bm\xi^{(J)},\mathbf{Z},\mathbf{Z^*}\right) ~\sim~ 
\mbox{Normal}\left(\frac{1}{|\mathcal{C}^*|}\sum_{k\in \mathcal{C}^*} \xi_{Z_k}\, ,\, \theta^{\xi} \right),
\]
The parameter $\sigma^*$ is sampled from a log-normal distribution with
\[
\mathbb{E}\left(\sigma^*\right) ~=~ \frac{1}{|\mathcal{C}^*|}\sum_{k\in \mathcal{C}^*} \sigma_{Z_k}
\qquad\mbox{and}\qquad
\mbox{Var}\left(\sigma^*\right) ~=~ [\exp(\theta^{\sigma})-1]\exp(2\mu^{\sigma} + \theta^{\sigma}).
\]

The proposal varies $\epsilon^*$ with respect to the current number $J$ of clusters. If $J>1$, $\epsilon^*$ is sampled with
\[
\epsilon^*~\sim~\mbox{Exponential}\left(\frac{|\mathcal{C}^*|}{\sum_{k\in\mathcal{C}^*}\epsilon_{Z_k}}\right).
\]
In case $J=1$, the model for $P_{k,k'}$~($k,k'=1,\ldots,K$) is fully described by the parameters $\gamma_1$ and $\beta$. If a second cluster is proposed, we have to sample two proposals to derive $(\gamma_0, \gamma_1, \gamma_2)$, and in order to satisfy the satisfy the dimension matching condition \citep{Green1995}. We sample $\epsilon_1^*$ and $\epsilon_2^*$ independently from an Exponential($\theta^{\epsilon}$) and define $\gamma_0^*$ as 
\[
\gamma_0^* ~=~ \gamma_1  \exp\left(\epsilon_1^*\right),
\quad\gamma_1^* ~=~\gamma_1, \quad \gamma_2^* ~=~ \gamma_1  \exp\left(\epsilon_1^* - \epsilon_2^*\right).
\]

We compute the acceptance probability as given in \citet{Green1995}. The determinant of the Jacobian is equal to 1, unless $J=1$, since the defined mapping is the identity function. For $J=1$, the determinant of the Jacobian is $\exp(\epsilon_1^*)$. For the reverse move, death, we select one of the existing cluster centreswith equal probability. The probability the accept the new cluster is thus
\[
\min\left\{1\, ,\,
\frac{L\left(\bm\theta^*_{\mathrm{M}},\bm\theta^*_{\mathrm{D}}\mid \mathcal{D}, \mathbf{Z^*}\right)}
{L\left(\bm\theta^{(J)}_{\mathrm{M}},\bm\theta^{(J)}_{\mathrm{D}}\mid \mathcal{D}, \mathbf{Z}\right)} \times \frac{\pi(\sigma^*,\xi^*,\epsilon^*)}{q(\sigma^*,\xi^*,\epsilon^*,\mid \bm\theta^{(J)}_{\mathrm{M}},\bm\theta^{(J)}_{\mathrm{D}},\mathbf{Z},\mathbf{Z}^*)} \times \frac{\kappa}{J} \times \frac{p_D}{p_B}
\right\},
\]
where $p_D$ and $p_B$ are the probabilities for proposing a death and birth move, respectively, $\pi$ is the joint prior density and $q$ is the joint proposal density. 

In case of a death, we first update the cluster labels to obtain $\mathbf{Z}^*$ for the proposed set of $J-1$ cluster centres. Let $(\sigma_*,\xi_*,\epsilon_*)$ denote the parameters of the cluster which is proposed to be removed, and $\bm\theta^{*}_{\mathrm{M}} = \bm\theta^{(J)}_{\mathrm{M}}\setminus\left(\sigma_*,\xi_*\right)$ and 
$\bm\theta^{*}_{\mathrm{D}} = \bm\theta^{(J)}_{\mathrm{D}}\setminus\left(\epsilon_*\right)$. The acceptance probability is then
\[
\min\left\{1~,~
\frac{L\left(\bm\theta^*_{\mathrm{M}},\bm\theta^*_{\mathrm{D}}\mid \mathcal{D}, \mathbf{Z^*}\right)}
{L\left(\bm\theta^{(J)}_{\mathrm{M}},\bm\theta^{(J)}_{\mathrm{D}}\mid \mathcal{D}, \mathbf{Z}\right)} \times \frac{q(\sigma_*,\xi_*,\epsilon_*\mid\bm\theta^{(J)}_{\mathrm{M}},\bm\theta^{(J)}_{\mathrm{D}},\mathbf{Z}^*, \mathbf{Z})}{\pi(\sigma_*,\xi_*,\epsilon_*)} \times \frac{J-1}{\kappa} \times \frac{p_B}{p_D}
\right\}.
\]

\subsection*{Remaining moves}
In case of a shift move, we first select one of the current cluster centres with equal probability and propose to reallocate it to one of the adjacent sites which is not currently a cluster centre. Let $j^*$ be the index of the sampled cluster centre. We first derive the set $\mathcal{N}$ of potential sites and then select one with equal probability as the new cluster centre $C^*$. This reallocation of a cluster centre usually changes the spatial clusters and we thus derive the set of updated cluster labels $\mathbf{Z}^*$ via \eqref{eq:Clustering}. To calculate the acceptance probability, we also require the set $\mathcal{N}^*$ of sites adjacent to $C^*$ which are not cluster centres; this set includes the current cluster centre $C_{j^*}$. Since $\bm\theta^{(J)}_{\mathrm{M}}$ and $\bm\theta^{(J)}_{\mathrm{D}}$ do not change, the prior densities cancel and the acceptance probability is
\[
\min\left\{1\, ,\,
\frac{L\left(\bm\theta^{(J)}_{\mathrm{M}},\bm\theta^{(J)}_{\mathrm{D}}\mid \mathcal{D}, \mathbf{Z^*}\right)}
{L\left(\bm\theta^{(J)}_{\mathrm{M}},\bm\theta^{(J)}_{\mathrm{D}}\mid \mathcal{D}, \mathbf{Z}\right)} \times \frac{|\mathcal{N}|}{|\mathcal{N}^*|}
\right\}.
\]

We specified three moves to update $\bm\theta^*_{\mathrm{M}}$ and $\bm\theta^*_{\mathrm{D}}$. Since the parameters for cluster $j$ and $j'$, $j\neq j'$, are independent given $\mathbf{Z}$, each cluster parameter is updated separately via an independence sampler; the proposal distribution is equal to the prior in this case and the acceptance probability is thus equal to the likelihood ratio. The last move updates the hyperparameters by drawing from the full conditional distributions using Gibbs sampling.

\bibliographystyle{apalike}
\bibliography{sample.bbl}

\end{document}